\documentclass[journal,twoside]{IEEEtran}

\usepackage{ifthen}
\usepackage{calc}
\usepackage{pifont}
\usepackage{color}
\usepackage{fancyhdr}
\usepackage{xspace}
\usepackage{xkeyval}
\usepackage{multirow}

\usepackage[normalem]{ulem}
\usepackage[english]{babel}
\usepackage{blindtext}

\usepackage{epsfig}
\usepackage{array}
\usepackage{algorithm2e}
\usepackage{amssymb}
\usepackage{amsfonts}
\usepackage{verbatim}
\usepackage{graphicx}
\usepackage{url}
\usepackage{amsmath}
\usepackage{xcolor}
\usepackage{hyperref}
\usepackage[justification=centering]{caption}
\usepackage{graphicx}
\usepackage{wrapfig}
\usepackage[numbers,sort&compress]{natbib}
\usepackage{enumitem}
\usepackage{algpseudocode}

\usepackage[T1]{fontenc}
\usepackage{hhline}
\usepackage{ctable}

\newcounter{hours}
\newcounter{minutes}

\newboolean{timeofmake}
\setboolean{timeofmake}{false}

\newcommand{\ignore}[1]{}

\newcommand{\dontinclude}[1]{ }

\newcommand{\putsec}[2]{\vspace{-0.05in}\section{#2}\label{sec:#1}\vspace{0.0in}}
\newcommand{\putsubsec}[2]{\vspace{-0.05in}\subsection{#2}\label{sec:#1}\vspace{0.0in}}

\newcommand{\tabput}[3]{
\begin{table}
\begin{center}
{
#2
}
\end{center}
\caption{#3 \label{tab:#1}}
\vspace{-0.20in}
\end{table}
}

\newcommand{\figput}[4][1.0\linewidth]{
\begin{figure}[t]
\begin{minipage}{\linewidth}
\footnotesize 
\begin{center}
\includegraphics[width=#1]{#2}
\end{center}
\vspace{-0.15in}
\caption{#4 \label{fig:#2}}
\end{minipage}
\vspace{-0.22in}
\end{figure}
}

\newcommand{\figputW}[4][\linewidth]{
\begin{figure*}
\begin{minipage}{\linewidth}
\footnotesize
\begin{center}
\includegraphics[width=#1]{#2}
\end{center}
\vspace{-0.2in}
\caption{#4 \label{fig:#2}}
\vspace{-0.20in}
\end{minipage}
\end{figure*}
}

\newcommand{\figref}[1]{Figure~\ref{fig:#1}}
\newcommand{\tabref}[1]{Table~\ref{tab:#1}}
\newcommand{\secref}[1]{Section~\ref{sec:#1}}

\newcommand{\name}{{Dart}\xspace}
\newcommand{\ack}{{\tt ACK}\xspace}
\newcommand{\ackplural}{{\tt ACKs}\xspace}

\usepackage{array}

\makeatletter
\newcommand{\thickhline}{%
    \noalign {\ifnum 0=`}\fi \hrule height 1pt
    \futurelet \reserved@a \@xhline
}
\newcolumntype{"}{@{\hskip\tabcolsep\vrule width 1pt\hskip\tabcolsep}}
\makeatother

\begin{document}

\title{\name: Divide and Specialize for Fast Response to Congestion in RDMA-based Datacenter Networks} 

\author{ 
Jiachen Xue, 
Muhammad Usama Chaudhry, 
Balajee Vamanan, 
T. N. Vijaykumar,
and Mithuna Thottethodi
\thanks{J. Xue was with the Department of Electrical and Computer Engineering, Purdue University, 
West Lafayette, IN 47907 USA.
He is currently with NVIDIA Corp., Santa Clara, CA 95051 USA 
(email: xuejiachen@gmail.com).}
\thanks{M. Chaudhry was with the Department of Computer Science, University of Illinois at 
Chicago, Chicago, IL 60607 USA. 
He is currently with VMware Inc., Palo Alto, CA 94304
 (email: chaudhryusama@gmail.com).}
\thanks{B. Vamanan is with the Department of Computer Science, University of Illinois at Chicago, 
Chicago, IL 60607 USA (email: bvamanan@uic.edu).}
\thanks{T. N. Vijaykumar, and M. Thottethodi are with the Department of Electrical and Computer Engineering, Purdue University, West Lafayette, IN 47907 USA (email: vijay@ecn.purdue.edu, mithuna@purdue.edu).}}

\markboth{IEEE/ACM TRANSACTIONS ON NETWORKING}%
{Xue \MakeLowercase{\textit{et al.}}: \name: Divide and Specialize for Fast Response to Congestion in RDMA-based Datacenter Networks}

\maketitle

\begin{abstract}
Though Remote Direct Memory Access (RDMA) promises to reduce datacenter network 
latencies significantly compared to TCP (e.g., 10x),   end-to-end congestion control in the presence of  incasts is a challenge. Targeting the full generality of the congestion problem,  previous schemes rely on slow, iterative convergence to the appropriate sending rates (e.g., TIMELY takes 50 RTTs). 
Several papers have shown that  even in oversubscribed datacenter networks most congestion occurs at the receiver. Accordingly, we propose a divide-and-specialize approach, called {\em \name},  which isolates the common case of receiver congestion and further subdivides the remaining in-network congestion into the simpler spatially-localized  and the harder spatially-dispersed cases. For receiver congestion, we propose {\em direct apportioning of sending rates (DASR)} in which  a receiver for $n$ senders directs each sender to cut its rate by a factor of $n$, converging  in only one RTT. For the spatially-localized case, \name provides fast (under one RTT) response by adding novel switch hardware  for {\em in-order flow deflection (IOFD)} because RDMA disallows packet reordering on which previous load balancing schemes rely.  For the uncommon spatially-dispersed case, \name falls back to DCQCN. 
Small-scale testbed measurements and at-scale simulations, respectively, show that 
\name achieves $60\%$ (2.5x) and  $79\%$ (4.8x) lower 
$99^{th}$-percentile latency, and similar and 
$58\%$ higher throughput than InfiniBand, and TIMELY and DCQCN.
\end{abstract}


\putsec{intro}{Introduction}

Many modern, interactive datacenter applications have tight latency 
requirements due to stringent service-level agreements 
(e.g., under 200 ms for {\em Web Search}). TCP-based datacenter networks 
significantly lengthen the application latency. 
Remote Direct Memory Access (RDMA)  substantially reduces latencies 
 compared to TCP by bypassing the operating system via hardware 
support at the network interface (e.g., RDMA over InfiniBand
and RDMA over Converged Ethernet (RoCE) can cut TCP's latency by 10x~\cite{
MPI-infiniband, MICA}). 
As such, RDMA may soon replace  TCP in 
datacenters~\cite{pilaf, FaRM, HERD, Panda-memcached}. 

Employing RDMA in datacenters, however, poses a challenge. 
RDMA provides hop-by-hop flow control 
and rate-based end-to-end congestion control~\cite{RDMA-IPDPS10,mellanox-whitepaper}. However,
RDMA's  congestion control is suboptimal for the well-known datacenter congestion problem, called {\em incast}, 
where multiple  flows collide at a switch causing queuing delays and long latency 
tails~\cite{dctcp} despite good network design~\cite{fat-tree-amin, fat-tree-original}.
Though  such congestion affects only a small fraction of the flows (e.g., $0.1$\%), 
datacenter applications' unique characteristics imply that  the  average latency is worsened.
For example, because Web Search aggregates replies from thousands of nodes, the 
$99.9^{th}$ percentile reply latency affects the average response time; or alternatively, dropping the slowest replies worsens the response quality. 
In TCP, incasts cause delays due to packet drops and 
re-transmissions~\cite{dctcp}. Though the lossless RDMA   does not incur packet drops, 
incast-induced queuing delays lengthen RDMA's latency tail~\cite{dcqcn}.

InfiniBand uses Early Congestion Notification (ECN) marks to infer imminent congestion
and cuts back the sending rates ~\cite{mellanox-whitepaper,RDMA-IPDPS10}.
While DCQCN~\cite{dcqcn} proposes a similar scheme for RoCE, TIMELY~\cite{timely} 
uses round-trip times (RTT) measurements, instead of ECN marks, for rate control in user-level TCP. 
Unfortunately, because ECN marks and RTT measurements need many round-trips 
to converge to the appropriate  sending rates (e.g., 50 RTTs in TIMELY), the schemes are too slow for the applications'
predominantly short flows each of which lasts only a handful of 
round-trips.  
During convergence, the schemes also lose  throughput due 
to over- and under-shooting the sending rates.

To speed up convergence, we leverage the result in several   papers~\cite{Benson2010NTC,eyeq,QiaoIMC,NDP} and reports from large datacenter operators such as Facebook~\cite{RoySIGCOMM2015}, Google~\cite{google-sigcomm15} and
Microsoft~\cite{KandulaIMC2009}: even under typical oversubscription most congestion in  datacenter networks occurs at the network edge
% ' top of the rack (ToR) switches ??? 
(i.e., at the link from top-of-rack (ToR) switch to the receiver) as opposed to within the network. Our simulations confirm this result which is due to high-bandwidth network core~\cite{fat-tree-amin, fat-tree-original}  and incast at the receiver. We make the key observation that while   general congestion  is complex and may require iterative convergence, the simpler and common  case of receiver congestion 
can be addressed quicker via specialization; 
{\em Without isolating this case, previous schemes  apply their iterative throttling to the general case. 
Instead, our proposal, called {\em \name}, employs a divide-and-specialize approach  to isolate receiver congestion and significantly speeds up the convergence.}
\name sub-divides the remaining case of in-network congestion into the simpler spatially-localized case  and the harder spatially-dispersed case. For the former where the network capacity is not under pressure (e.g., due to imperfect ECMP hashing), \name avoids throttling which is unnecessary. 
For the latter where the network capacity is under pressure (e.g., due to dynamic network load spikes), \name falls back on DCQCN's throttling which may be unavoidable.
Load balancing~\cite{MPTCPSIGCOMM11,spain,pktscatterinfocom,detail,flowbender,Presto16,CONGA,dibs} can alleviate localized in-network congestion but not receiver congestion, and usually reorders  packets which is not supported by RDMA.

To address  receiver congestion,  
we make the key observation that unlike in a wide-area setting, 
datacenter applications are co-operative 
where  a receiver of $n$ senders can direct each sender to cut
its rate by a factor of $n$,
This mechanism, called {\em direct apportioning of sending rates (DASR)}, ensures that the critical, short flows get their fair share of  (instantaneous) throughput without being swamped 
by the background, long flows.
When a sender completes, the (instantaneous) sending rate is adjusted as per the 
new sender count. Because  DASR piggybacks the count in the receiver's acknowledgments  to the 
senders, DASR achieves accurate and  {\em one-RTT} convergence of sending 
rates without any repeated adjustments, unlike  previous schemes. 
Specifically,  (1) RCP~\cite{rcp} proposes to apportion the rates among the senders, but employs slow, iterative convergence {\em at the switches}  {\em because RCP (a) targets general congestion  without isolating receiver congestion and 
(b) uses general parameters to arrive indirectly at fair share instead of directly counting flows which is hard to do at Internet scales};  we evaluate RCP's convergence in ~\secref{rcp}. (2) EyeQ~\cite{eyeq} highlights edge congestion but  applies  RCP's iterative convergence, which takes 25-30 RTTs, without specializing for edge congestion.  (3) NUMFabric~\cite{numfabric} achieves more flexible and  faster bandwidth allocation than TCP but still employs iterative convergence (e.g., 31 RTTs). And, (4) 
while ExpressPass~\cite{expresspass} and  NDP~\cite{NDP}  target general congestion via receiver-based congestion control,
neither scheme isolates receiver congestion.
ExpressPass employs BIC-TCP iterative convergence  which takes 20 RTTs for a datacenter network  (\secref{rcp}); ExpressPass shows results only for a simple network. 
NDP fundamentally relies on  (a) packet spraying, which reorders packets, to reduce congestion  and (b) packet trimming, which removes payloads, to unclog congestion notification to the receiver. Neither of these  mechanisms is  supported by RDMA which has no software stack like TCP. 
Without these mechanisms, NDP would see  more congestion and  slower feedback.  
DASR's faster convergence reduces 
latency tail (critical flows quickly get their share) and improves throughput (fewer adjustments). 
In an additional optimization, DASR leverages application-provided incast degree to avoid counting the senders and converge even faster.

To address spatially-localized, in-network congestion, 
\name  simply deflects the affected packets  under the premise that 
an alternate path is  faster than being queued up in the shortest 
path.  To avoid livelock,   \name allows only a 
few deflections for a  packet after which the packet is not 
deflected even at a congested switch. \name avoids deadlocks via  a widely-used virtual-channel-based scheme~\cite{Duato:original,mellanox-whitepaper}.  
Because RDMA does not support packet reordering, \name provides hardware support in the switch to keep a flow's 
packets in order.  While deflection~\cite{hot-potato}  is  well known, our 
contribution is {\em in-order flow deflection (IOFD)} unlike previous load-balancing schemes including DIBS~\cite{dibs}. 
As a congestion response, deflection is much lighter-weight and quicker (well under one RTT) than
 rate-cutting using iterative convergence and does not affect the sending rates.
For spatially-dispersed in-network  congestion, which is uncommon, \name falls back to DCQCN's heavy-weight rate modulation.   By filtering out receiver congestion and localized in-network congestion, \name cuts the number of ECN marks, which trigger DCQCN fall-backs,  by 4x for typical workloads.  

We make  four observations: 
First, receiver congestion is easy to differentiate from in-network congestion (\secref{nonreceiver}).
Second, DASR works only for receiver congestion but not  for in-network congestion (e.g., two flows collide in the network but go to different receivers which cannot detect the collision); and vice versa for IOFD (flows colliding at the receiver should not be deflected). 
As such, one of our contributions is identifying the specific case and applying the appropriate specialization. 
Third, because DASR and IOFD {\em separately} target receiver congestion and localized in-network congestion, respectively, they are more effective despite  being  simpler than previous schemes which tackle the full generality of the problem using a common mechanism.  Finally, \name leverages RDMA's unique features. While 
DASR is applicable to both RDMA and TCP, our DASR implementation 
relies on RDMA's discrete messages as opposed to TCP's continuous flows (~\secref{shortflows}).  IOFD  specifically addresses RDMA's lack of support for packet reordering, 

In summary, our  key contributions are:

\noindent
$\bullet$
employing a divide-and-specialize approach to congestion control;

\noindent
$\bullet$ addressing 
receiver congestion via 
{\em direct apportioning of sending rates}   by using the sender count
to achieve   accurate and faster, one-RTT convergence of sending rates 
than previous schemes which are iterative; and

\noindent
$\bullet$ 
addressing
spatially-localized in-network congestion via {\em  in-order flow
deflection} whereas previous schemes reorder packets which is not supported by RDMA.

A small-scale 16-node testbed implementation  shows that \name converges to the desired sending rate in one RTT and achieves $60\%$ (2.5x) lower latency than and similar throughput as  InfiniBand. 
Datacenter-scale {\em ns-3} simulations show that \name achieves  $79\%$ (4.8x) lower 
$99^{th}$-percentile latency and
 $58\%$ higher throughput, on average,  than TIMELY and DCQCN for  typical  over-subscription and load settings.

\putsec{background}{Challenges and  Opportunities}

Modern datacenter applications demand \textit{both} low latency tails and 
high throughput from the network. Interactive datacenter applications, such 
as \textit{Web Search}, generate thousands of short flows to lookup large 
distributed datasets for \textit{each} user query. 
As described in~\secref{intro}, the overall response time  is bound by the  
$99^{th}$ - $99.9^{th}$ percentile of flow completion times 
(i.e., the tail-latency problem)~\cite{tail-cacm}.
 Further, the synchronous 
nature of the lookup responses, which are aggregated in subsets, implies 
that each subset arrives at a switch causing an {\em incast}, which worsens when multiple queries' subsets arrive at the same time.  On the other 
hand, background applications  (e.g., Web Index update)
demand high 
throughput for  large volumes of Internet data.
These long flows colliding with the short flows also exacerbate incasts.

The OS overheads in TCP drastically dilate 
network tail latencies (e.g., $99^{th}$ percentile latency is 10-20x of 
median latency~\cite{dctcp}). Further, a slow response to congestion  hurts latency at the start of incasts and throughput at the end.  
Similarly, an inaccurate response  affects 
latency or throughput, depending on whether the rate was less or more 
than the optimum. 

\putsubsec{RDMA}{RDMA}

With RDMA, the application invokes the NIC directly without involving 
the OS -- (1) At the sender, the NIC uses 
DMA to copy data from the application memory to its buffers using DMA 
and sends the data after some protocol processing; (2) At 
the receiver, the NIC copies data into the receiving application's buffer. 
Thus, RDMA eliminates OS intervention and accelerates protocol processing 
at both the sender and the receiver. The buffers are pinned in physical 
memory 
and the address translations are cached at the NIC during connection 
establishment. RDMA-based transports \cite{FaRM, HERD} 
show an order-of-magnitude reduction in flow latencies at low 
loads. As such, RDMA, initially proposed for multiprocessor 
networks \cite{VIA}, is finding its way into modern datacenters.

\putsubsec{challenges}{Challenges}
Existing RDMA transports provide hop-by-hop flow control to ensure lossless operation. For 
example, InfiniBand \cite{ib} employs credit-based flow 
control and RoCE \cite{roce} uses Priority-based Flow 
Control (PFC). InfiniBand provides rate-based end-to-end congestion
control using ECN  marks~\cite{RDMA-IPDPS10,mellanox-whitepaper}.
DCQCN \cite{dcqcn} has shown that RoCE without end
-to-end congestion control degrades in both latency and throughput at 
high loads.

As discussed in~\secref{intro}, previous schemes address  the full generality  of the congestion problem and end up with iterative convergence to the appropriate sending rate upon congestion. Unlike TCP's window-based rate control,   RCP's~\cite{rcp}  routers iteratively calculate and  convey the fair-share bandwidth to the senders sharing a link, which slows convergence (see \secref{rcp}). 
DCQCN \cite{dcqcn} and TIMELY \cite{timely} improve end-to-end congestion control  at datacenter scales for  RDMA (RoCE) and user-level TCP respectively. Both DCQCN and TIMELY
directly control the sending rate by pacing the packets sent out of the NIC.   DCQCN starts a flow at the full line rate, employs  ECN marks as feedback and cuts the sending rate in proportion to the exponentially-averaged fraction of ECN-marked packets. To avoid some problems of ECN (e.g., low-priority packets may not see ECN marks), 
TIMELY  employs RTT measurements as feedback and modulates the sending rate (additive increase and multiplicative decrease) based on RTT gradients bounded by thresholds at the extremes. 

Despite these innovative ideas, because these schemes tackle the general case with arbitrarily changing number of flows which interact in arbitrary ways,  the schemes rely on slow, iterative convergence to the appropriate sending rates. As discussed in~\secref{intro}, other schemes, including EyeQ~\cite{eyeq}, NumFabric~\cite{numfabric} and ExpressPass~\cite{expresspass}, also rely on iterative convergence. Such convergence  requires many round trips (e.g.,  60 RTTs in RCP, 50 RTTs in TIMELY, 31 RTTs in NUMFabric, and 25-30 RTTs in EyeQ), as illustrated  
in \figref{convergence} for a sender whose initial sending rate is 100\%
of the line rate and the target rate is 50\%. The upper half of 
\figref{convergence} shows the  tuning of sender-inferred rates. Such iterative convergence hurts both latency and throughput, as we show in \secref{main}. Because DCQCN and TIMELY specifically target 
RDMA (RoCE) and user-level TCP (which bypasses the OS like RDMA),  respectively,  and are representative of iterative convergence, we compare \name to these two schemes in our results. 

\putsubsec{opportunity}{Opportunities}

\figput{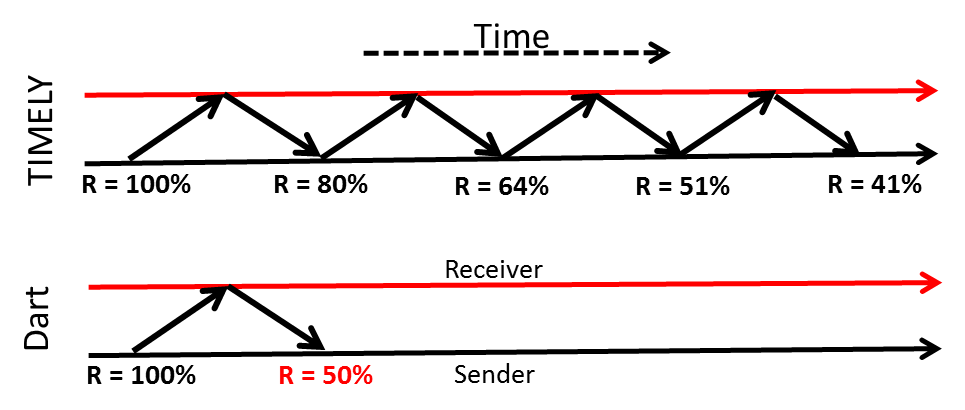}{}{\name's fast, one-RTT convergence}

\name employs a divide-and-specialize approach to avoid iterative convergence in the common case of receiver congestion (i.e., multiple senders intentionally sending to a receiver). For this case, \name uses {\em direct apportioning of sending rates (DASR)} which specifies  the appropriate  sending 
rate in one RTT without repeated adjustments (see the lower half of \figref{convergence}).  Thus, \name achieves accurate and 
fast convergence for receiver congestion. \name further sub-divides the remaining case of in-network congestion into two sub-cases:  the easier spatially-localized  congestion and  the harder spatially-dispersed congestion. For the localized sub-case, \name employs in-order flow deflection (IOFD) which does not affect the sending rates. 
Such deflection  is a quicker, lighter-weight, in-network response (well under one RTT)
than the previous schemes' iterative convergence. For the dispersed sub-case,  which is uncommon especially after IOFD filters out localized congestion,  \name falls back to DCQCN.

\putsec{throughput}{Receiver Congestion}
 
We  start with  direct apportioning of sending rates  (DASR) and describe in-order flow deflection (IOFD)  in \secref{latency}. We note that when contention is at the end-points, the fair share of 
bandwidth for each of $n$ (say) senders is well-defined as $\tfrac{1}{n}$. The fair share
can be extended easily to weighted fair shares.

In our description of DASR and IOFD, we use the term 'flows' to mean RDMA messages. Short flows are effectively small messages 
(e.g., those that contain small search queries for web-search, or key-value lookup requests for memcached). Long flows are effectively large messages that  perform bulk-copying of large sections of memory (e.g., for index-updates in web-search). 
Both short and long flows may be packetized as necessary. While flow sizes are not known to the TCP layer,  message sizes must be sent
explicitly in RDMA and hence the RDMA application messaging layer can identify long and short flows.

\figput{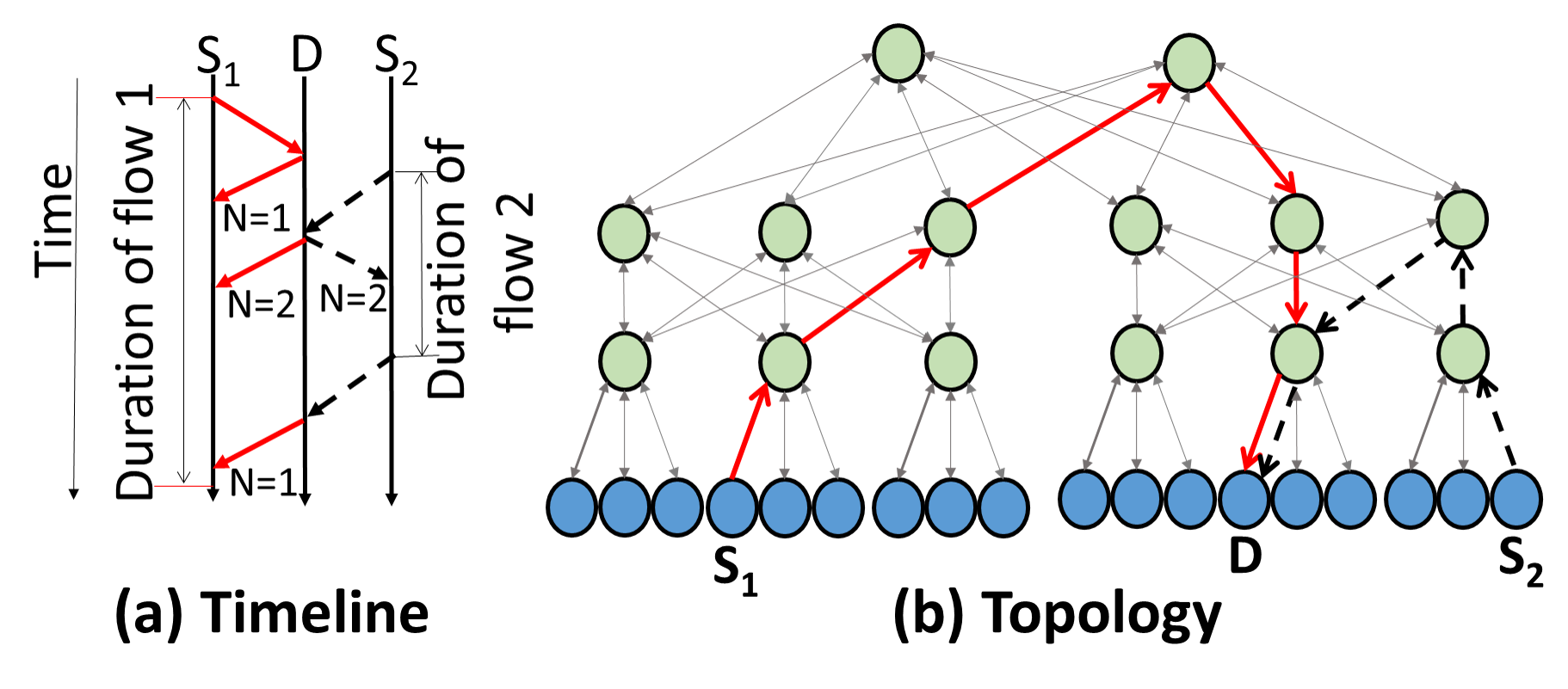}{}{Direct apportioning of sending rates}

\putsubsec{operation}{Direct apportioning of sending rates}
All flows begin at the full line rate because (1) we want to avoid penalizing the latency of short flows, 
and (2) \name's fast feedback can quickly throttle long flows if necessary. 
\name piggybacks the sender count, the $n$ value,  with \ackplural to all the senders; \ackplural use
high-priority queues in \name as well as all the other schemes we compare.
Such piggybacking can be achieved via NIC firmware without hardware changes. (In practice, implementing firmware changes on proprietary NICs is not feasible without vendor support. We discuss our prototyping approach later in \secref{testbed}.)
As such, senders receive continuous, fast -- one-RTT --  direction from the receivers
 on their allowed transmission rate. Such co-ordination is between the 
end-point NICs; the switches need not be modified. 

\figref{largeflows} shows an oversubscribed fat tree to illustrate \name's  operation in terms of fair-sharing among long flows.
Consider the example shown in \figref{largeflows}(a) wherein
a single receiver ($D$)  receives a steady long flow from one sender ($S_1$)
at the line rate. That sender continues to transmit at the line rate without 
throttling as it sees the $n$ value remain $1$ in the \ackplural from the
receiver. 
When a second sender ($S_2$) initiates another long flow to the same receiver  ($D$), 
there is contention at the leaf-level switch, as shown in~\figref{largeflows}(b)
where the solid and broken lines show the two flows.  As the two flows' packets
arrive interleaved at the destination node, the receiver's NIC piggy-backs the updated $n=2$ value 
with the \ackplural to each sender. The \ackplural cause the sender NICs to throttle the 
rate to $\frac{1}{n} = \frac{1}{2}$ of the line rate, which can be sustained in steady state.

The above discussion illustrates the two key benefits of 
DASR. 
First, the continuous feedback mechanism means that congestion
control feedback to senders is fast, in one RTT. Second, the senders are given 
an accurate and precise rate not to 
exceed. The algorithm seamlessly handles flow ``churn'' by constantly
sending updated $n$ values.  

\putsubsec{shortflows}{Short flows and incasts under DASR}

The case of short flows, including incasts, interacting with long flows uses the same mechanism
to ensure that the latency of short-flows is not hurt (\figref{shortflows}(a)).  
A long flow
that contends with $k$ other short flows from $k$ unique senders is directed to reduce its sending rate
to $\frac{1}{k+1}$ because $n=k+1$. While this
throttling helps the short flows' latency,
such throttling is short-lived and does not hurt the long flow's throughput. The presence of short flows
can be treated as a case of flow-churn; the long flows throttle their rates
according to the number of short flows, but only for the duration of the short flows (\figref{shortflows}(b).

The rate throttling at the sender is staggered 
by the time required for the receiver's \ack (with the piggy-backed $n$ value) to reach the sender.
while DCQCN and TIMELY also incur this \ack delay (\secref{background}),
the previous schemes require several iterations  of
RTT measurements or ECN marks, involving several round trips,  for the
sender to infer the appropriate  rate (e.g., 50 RTTs in TIMELY).
This delay hurts both short flows' latency and long flows' throughput.
In contrast, DASR converges in one RTT to the
appropriate sending rates. 

\figput{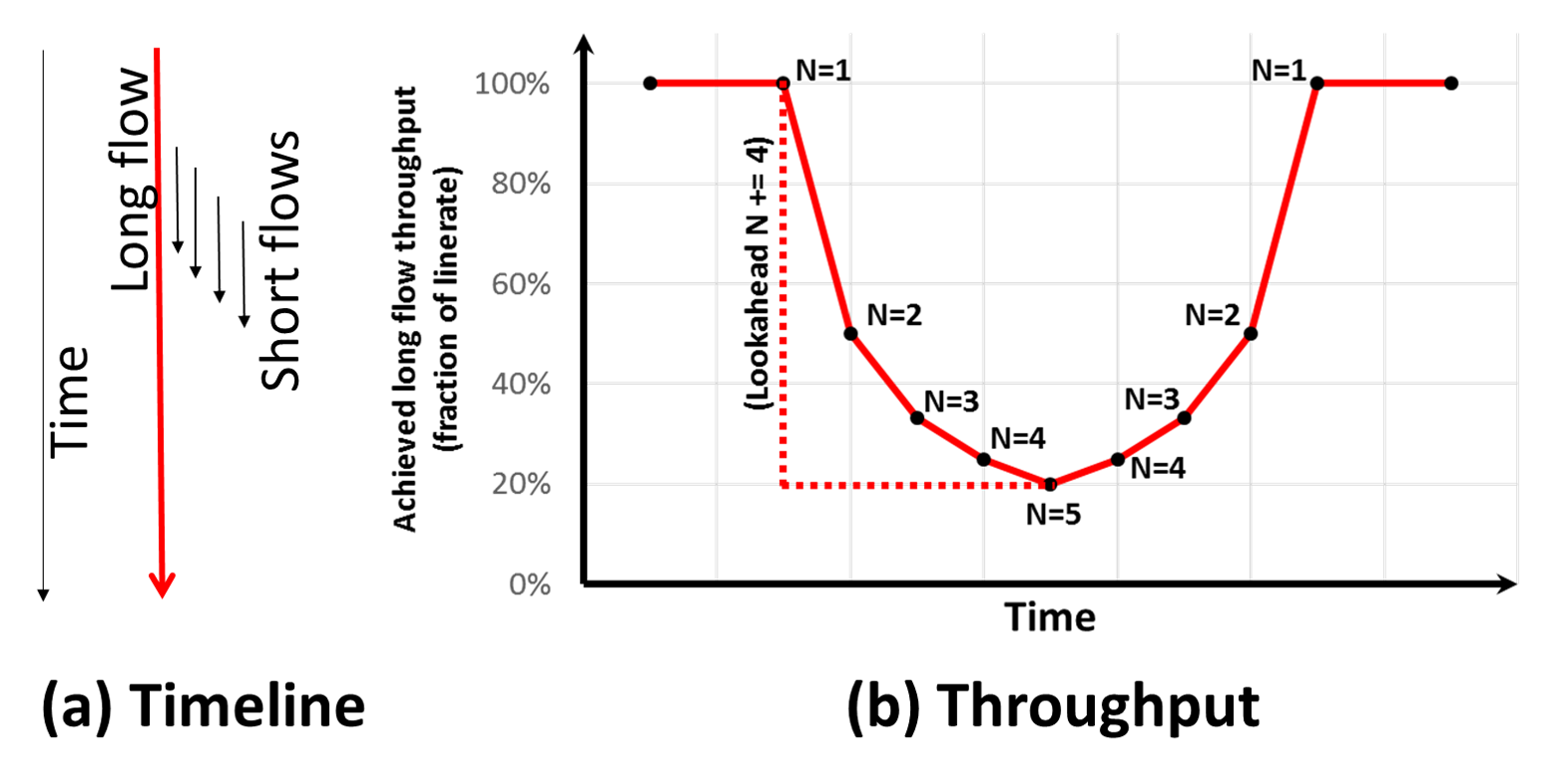}{}{Short flows mixed with long flows}

\name addresses one other challenge: accurate counting of senders. 
Consider a case where two incasts to the same destination (say $D$) 
begin close in time and there is an overlap
in the senders of the two incasts (sender $S$ is part of both
incast groups).  Because $S$'s two incast flows would be serialized at $S$'s NIC, $D$'s NIC should count source $S$ exactly once when determining $n$.
This case is handled naturally because \name tracks {\em in software} the unique senders  
of active flows -- in the Active Unique Sender Set (AUSS).
Upon a new message/flow, the sender of the message is added to the 
AUSS if not already present (see \figref{flowchart}(a)). 
Further, \name initializes a count of in-flight messages associated with that sender
to 1 (if not previously present in the AUSS) or increment the in-flight message count (if previously present in the AUSS and  multiple messages from the same sender are concurrently
active). \name finally decrements the sender count only when all the messages from that sender terminate, as shown in \figref{flowchart}(b). 
With the above tracking in place, DASR can use the 
number of elements in the AUSS as the $n$ value (i.e., $n = |AUSS|$).

Finally, each sender in the AUSS is associated with a timestamp of
the flow's last packet. Any flow that is idle for long (e.g., 2 seconds) is assumed to be dead and eliminated from the AUSS. This well-known soft-state
approach ensures that DASR does not artificially throttle active senders in cases where other senders may fail after initiating message transmission. 
% Such flows eventually timeout and are eliminated from the  AFS. 
Recall that RCP requires switch support to handle the full generality. In contrast, \name requires extra state only at the receiver (host) to specialize the common case of receiver congestion. 

RDMA's connectionless nature (unlike TCP) and its clearly-marked message start/end ensures that senders are  not counted
in  idle periods (as shown in \figref{flowchart}). 
Because our DASR implementation relies on RDMA's message start/end markers for accurate AUSS tracking, it does not extend to TCP  which views communication as a continuous stream without markers making it hard to account for flow idleness.  

\figput{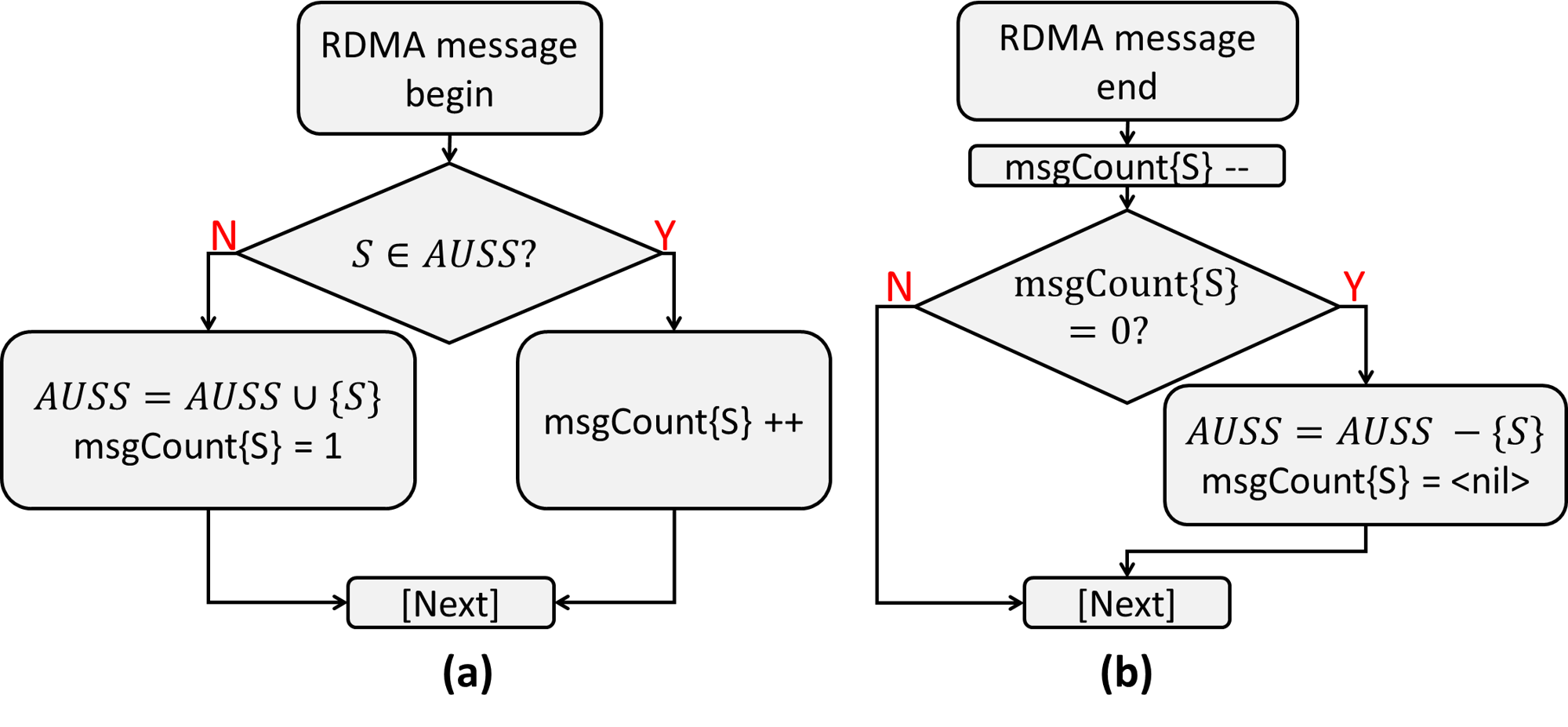}{}{Active Unique Sender Set for sender S (in software)}

\putsubsec{nonreceiver}{Handling non-receiver congestion}

\figref{dasrfsm} illustrates our state machine that \emph{exhaustively} handles receiver and non-receiver (in-network and at source) congestion. 
\name distinguishes between receiver and non-receiver congestion based on  two observable symptoms: (1) throughput at the receiver, and (2) ECN marks. Changes in either of the two 
trigger state changes as shown in \figref{dasrfsm}.

As long as no ECN marks are received, \name remains in the ``No Congestion'' state. 
While DASR targets receiver congestion,  
both receiver congestion and non-receiver congestion (including network and source congestion at the sender's NIC)
may result in ECN marks. 
For source congestion, we require that
the source NICs be capable of ECN marking, which is possible in today's SmartNICs. 
For example, we can programmatically set ECN on Netronome Agilio CX NICs 
based on queue depth, which is accessible as intrinsic metadata~\cite{agilio}.
Without any additional safeguards, the ECN-based DCQCN fall-back may over-throttle the sending rates in addition to DASR even  for receiver congestion. 
To avoid such over-throttling, we observe that during receiver congestion, the throughput seen by the receiver is not affected as all flows headed to that receiver would be serialized anyway at the
last hop (i.e., the receive throughput is equal to line rate). This condition triggers DASR, denoted by ``Receiver Congestion'' state in \figref{dasrfsm}. In this state, \name piggy-backs the $n$ values while suppressing the ECN marks on the returning \ackplural. 
The throughput is unaffected even if receiver congestion occurs at 
an internal switch  -- it is still receiver congestion
irrespective of where it occurs.

In contrast, in the case of non-receiver congestion (i.e., network and source congestion) where contending flows are headed to different destinations, the bottleneck link capacity would be shared by contending flows. As a result, when the flows eventually reach their destinations, the receivers would observe throughputs that are less than the line rate. In addition, the receiver would also observe ECN marks due to congestion. Accordingly, \name enters ``Non Receiver Congestion'' state when the receiver observes lower than line rate as well as ECN marks. Because DASR cannot handle non-receiver congestion, the receiver allows ECN marks, which trigger DCQCN at the senders. Finally, to avoid DASR from interfering with DCQCN, \name sets $n=1$ in this state. 
While the above description handles receiver and non-receiver congestion occurring separately,  
\name naturally handles the case of the two together in two steps.
In the first step, \name enters the ``Receiver Congestion'' state causing DASR to kick in. For non-receiver congestion, however, DASR's apportioning may cause the senders to underutilize their throughput share. In that case, the receiver rate would fall below
the line rate, causing a transition to the ``Non-Receiver Congestion'' state in the second step, where DSQCN kicks in to avoid continued throughput loss. 
Thus, \name \emph{exhaustively} covers all cases of congestion among the three states in \figref{dasrfsm}.

\paragraph{\textbf{\name's convergence:}}
From \figref{dasrfsm}, it is clear that DASR covers only the special case of receiver congestion and converges to the \emph{correct} sender rate (i.e., fair share). During non-receiver congestion (i.e., in-network or source congestion), \name falls back to DCQCN. \name's convergence is thus guaranteed by DCQCN's convergence in this case. Overall, because our state machine exhaustively covers \emph{all} congestion states, \name converges to the correct sender rates in \emph{all} cases.

\figput[1.0\linewidth]{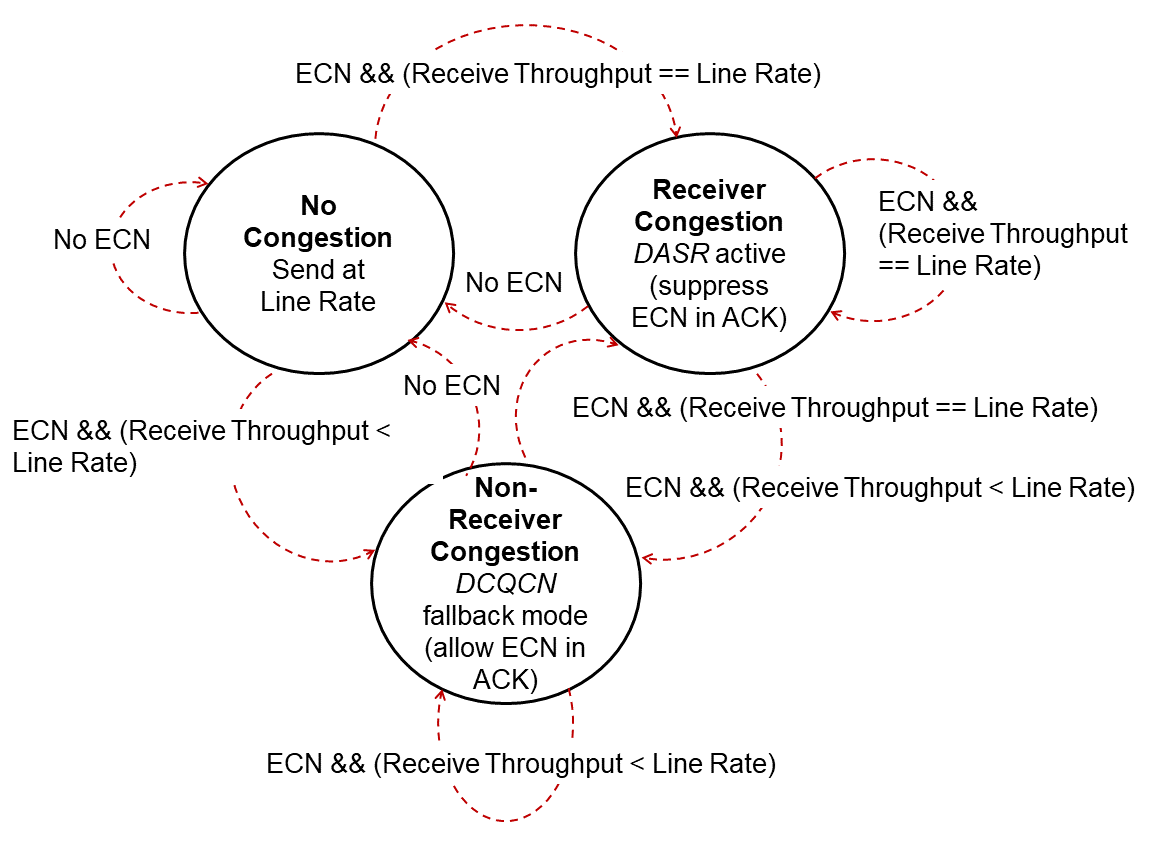}{}{Handling receiver and non-receiver congestion}

\putsubsec{lookahead}{Accelerated DASR}
We further improve \name's performance by having the application 
provide a {\em look-ahead} notification of the upcoming set of incast flows that are part of an incast group. For example, 
if each incast message carries (1) information that it is part of a 20-flow incast and (2) the list of the 20 senders, 
the receiver NIC can advertise rate limits to the 20 senders
after just the first such message, even 
before the other senders' packets arrive at the receiver. 
As with the $n$ value, such lookahead notification can also be handled via NIC firmware.
Thus, 
the AUSS can be populated with the set of senders in advance of actual packet arrival from all the senders. The long flows back off quicker with
this look-ahead, as shown by the dotted line in \figref{shortflows}(b).
For accurate counting, \name treats any flow as if it begins when the look-ahead
notification first arrives. The ending of flows is handled  as without the look-ahead. The look-ahead overhead is reasonable (e.g., 20 two-byte sender-ids, each of which can address 64K sender NICs, amount to 40-byte or 2\% overhead for a 2-KB payload). 
Unlike  generic applications, latency-sensitive applications are specialized where the incast groups -- {\em static} in the application --  are likely known to the programmer (e.g., Web Search). Identifying the static groups is enough even if they {\em dynamically} and unpredictably break into subsets at different switches because eventually the whole static group causes receiver congestion which is DASR's target.

\putsubsec{rdrcfailures}{Failures and attacks under DASR}
Because the AUSS tracking uses soft-state (as described in \secref{shortflows}), \name can handle failures seamlessly. Any flow in 
the AUSS (irrespective of whether it uses look-ahead) will naturally timeout and exit the AUSS when senders fail.
However, untrusted entities in multi-tenant datacenters may attempt denial-of-service attacks by frequently sending look-ahead notifications which results in other senders throttling themselves. To ensure SLA compliance, datacenters typically 
use rate-limiting to ensure that VMs of a tenant do not exceed their fair share of bandwidth. \name's AUSS tracking can be private to 
individual tenant's flows. As such, any false information from one tenant 
can not affect other tenants' flows. As a last resort, the look-ahead optimization can be turned off in multi-tenant datacenters, while retaining the main DASR which is not susceptible to such attacks. We isolate the look-ahead's performance from that of the main DASR in~\secref{isolate}. 

\putsec{latency}{Localized In-network congestion}
We now address in-network congestion, starting with the easier spatially-localized congestion, including incasts,  and then discuss the harder spatially-dispersed congestion.
Localized in-network  contention is usually the result of 
temporary link contention in a small neighborhood of switches. Such contention 
may result in packets being unnecessarily serialized (e.g., even though 
they may be headed to different destinations).
In such situations, \name deflects {\em all} the packets of selected short-flows to avoid this serialization penalty. 
Consider the example shown in \figref{misroute} with two flows
between the source-destination pairs $(S_1,D_1)$ (solid arrows) and  
$(S_2,D_2)$ (dashed arrows). 
Assuming the second flow (dashed arrows) finds one of the links congested, 
the flow may take an alternate path, away from the congested link -- a response well under one RTT. While such deflection results in additional hops (two in the example -- one misroute and another to recover from the misroute), \name's deflection policies ensure that  (1) this penalty is far lower
than that of the serialization so that deflection significantly improves latency over previous schemes' iterative convergence, and (2)  the relative overhead of extra link utilization is low   (\secref{policies}). Further, our design
is free from livelocks and deadlocks (\secref{policies}). 
We describe below \name's mechanisms 
and policies for  such deflection-based congestion avoidance.

\figput[0.9\linewidth]{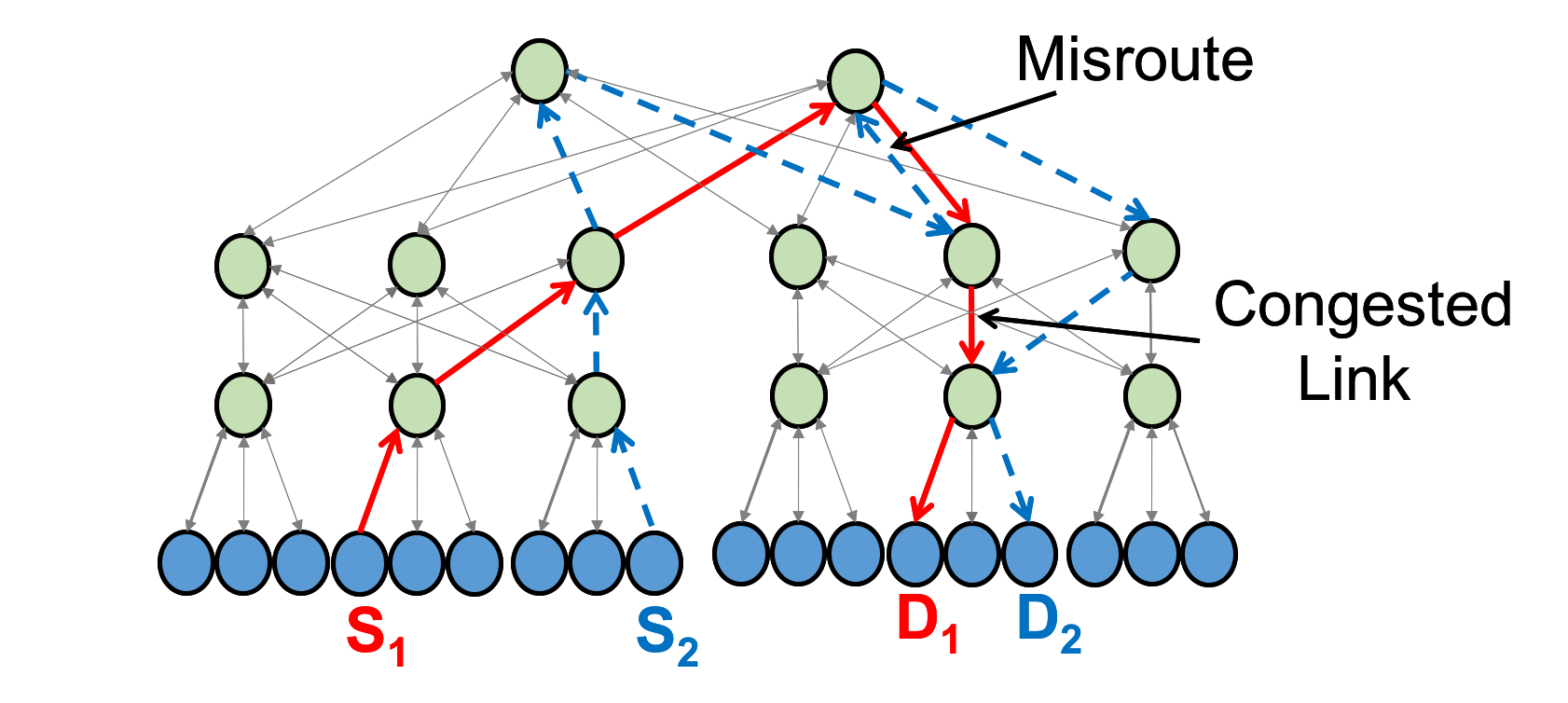}{}{Misrouting to avoid congested links}

\putsubsec{mechanisms}{In-order flow deflection mechanisms}

Deflection routing is a well-known technique for load balancing~\cite{hot-potato}.
In general, deflection routing can cause reordering of packets. As such, 
deflection is relatively straightforward to use when either the application 
does not require ordered packet delivery or there is a reassembly layer that 
reorders received packets to be delivered in the correct order (e.g., TCP). 
Indeed, in addition to being well-explored in other contexts, 
such packet-by-packet deflection has also been proposed for congestion avoidance in data centers (DIBS~\cite{dibs}).

In contrast, for RDMA networks, there is no software stack to reassemble 
out-of-order packets of a message/flow. 
Consequently, the limited hardware support for packet reordering are cases 
where re-ordering does not change the semantics. For example, 
current support in 
ConnectX5~\cite{connectx5} is limited to read/write RDMA verbs. A bulk remote-write 
can be broken into many pieces and the order among the pieces is not important as 
long as they all complete.  However, such reordering can change the semantics 
in send/recv based RDMA verbs. For example, a flow abstraction, if broken up into pieces, needs (1) sequence numbers associated with each piece, and (2) reassembly 
at the receiver to put the pieces back together in sequence order.  Note that 
send/recv verbs  are widely used {\em and} acknowledged as higher performing than 
read/write for server applications~\cite{HERD}.  
Recent work by Mittal {\em et al}~\cite{mittal-rdma} extends modest support even for reordering of flows with such sequence numbers; but with a fixed hardware window for reassembly. Such limited window size requires the source to throttle packets to ensure that the sliding window does not overflow, which reduces throughput. 
On the other hand, packet-by-packet deflection would require the ability to handle unbounded reordering (and not just the limited reordering support in ~\cite{mittal-rdma}), which can impose significant CPU overheads~\cite{Presto16}.

To avoid such overheads, the network must guarantee
in-order delivery semantics. For such networks, \name uses novel 
{\em in-order flow deflection (IOFD)} instead of the above packet-level deflection. 
The key challenge in IOFD is to ensure that later packets
of the flow traverse the same network path as the header packet 
of the flow. Further, the semantics do not allow for any 
false-positives (i.e., the switch misidentifies a non-deflected flow
as a deflected flow) or false negatives (i.e., the switch `forgets' a
misrouted flow to be one). Such strong semantics may seem challenging
especially when considering router failures. We describe the fault-free case below and  address faults in~\secref{faulttolerance}.

A naive solution would be to maintain routing history 
in the switch for every flow which may be many at a given time, and look up the history for  every packet.
Fortunately, because only short flows are latency-critical, IOFD applies only to short flows
only a few of which overlap at a switch at any given time  (say 4 to 8).
Long flows that collide at the receiver are handled by DASR.
Some spatially-dispersed in-network congestion due to long flows
is inevitable despite best-in-class hashing and other schemes~\cite{MPTCPSIGCOMM11,flowbender,Presto16}. 
% of long flows over the core links). 
In our design, such collisions trigger the  DCQCN fall-back. 
Crucially, the latency-critical short flows are deflected away from 
such  collisions.  Recall that flow sizes are known in RDMA (\secref{throughput}).

IOFD  maintains the set of misrouted flows in a small 
content-addressable memory (CAM) called the {\em deflected flow table (DFT)} at each router. 
Entries in the DFT are allocated when the start packet of an 
RDMA message  is 
chosen for deflection {\em and} a free entry is available in the DFT.
Each entry includes the flow id or RDMA message id (the searchable field) 
and a randomly-selected output port for that flow 
(the data field of the table entry).
Entries in the DFT are de-allocated when the end packet of an RDMA message  passes through the switch.
To ensure that the history of misrouted flows is not lost, no DFT entry
may be overwritten except by natural deallocation. To avoid livelocks,
IOFD allows a packet only a limited number of {\em misroutes} 
which are encoded as {\em deflection token} bits in each packet header. 
The switch removes a token from each misrouted packet.
If either DFT entries are unavailable or the flow has exhausted its 
tokens, the flow may not be deflected at the switch.
Every packet consults the DFT  to determine  its path, as shown in
\figref{dft}. If the packet's flow-identifier matches one
of the entries, the packet is deflected to the port indicated by 
the entry (e.g., a packet matching DFT entry id = 0xBC is deflected to port 12 in \figref{dft}).
To ensure that messages do not end up at unintended end-nodes,
leaf-level switches (ToR switches) deflect messages back to the network and not to end-nodes.
Note that, because the DFT is small (e.g., 8-entry CAM), 
the delay and power overheads are negligible.

\putsubsec{policies}{IOFD policies}

There are three policy decisions that IOFD makes to strike a balance
between over-aggressive deflection and inadequate deflection.
First, to determine if deflection is competitive (i.e., the expected 
queuing delay at the switch is high enough that a few 
additional network hops may  be better), IOFD compares the current queue position to an
empirically-determined {\em deflection threshold}. Deflection is allowed only if 
the queue position is above the threshold.
Second, to avoid unnecessarily-long deflection chains and livelock caused by loops,
IOFD deflects only the packets with spare deflection 
tokens (\secref{mechanisms}). Once the deflection tokens are exhausted, 
a packet incurs the full latency penalty of waiting in the network queues. 
Finally, the possibility of deadlocks must be carefully handled. 
Specifically, modern DC networks 
(Clos variants~\cite{clos}) typically use Valley-free routing~\cite{valleyfree} or up*/down* routing~\cite{duato97}
to guarantee deadlock-freedom. 
Although IOFD can violate the rules of valley-free-routing, such violations are possible independent of IOFD. Hu {\em et al}~\cite{tagger} show the violations of valley-free routing in real data-center measurements for a RoCE network. 
As such, IOFD can leverage the same (or similar) mechanisms that
are used to handle failures to handle flow deflections. 
We employ one such well-understood {\em deadlock avoidance}~\cite{Duato:original} technique by leveraging virtual channels (`virtual lanes' (VL) in InfiniBand~\cite{mellanox-whitepaper}).
Deadlock avoidance employs two class of VLs: (1) escape VLs that are guaranteed to  avoid cyclic buffer dependencies and (2) non-escape VLs that may incur cyclic buffer dependencies.  Packets/flows may move from non-escape VLs to escape VLs (which ensures that flows in non-escape VLs can always make forward progress) 
but {\em not} vice versa. 
In our context, if traffic on one virtual lane (VL) -- the escape VL -- is not deflected, and  flows that traverse the escape VLs never flow back to non-escape VLs, deadlock-freedom is guaranteed.
Unlike  {\em deadlock prevention} which places routing restrictions that avoid certain turns (e.g., ~\cite{Dally-deadlocks,valleyfree}), deadlock avoidance works without preventing any turns~\cite{Duato:original,duato97}; rather it takes the approach that any turn may be allowed by at least some VLs. 
Recent work~\cite{DCdeadlocks} discusses  deadlocks, {\em other} than routing deadlocks, created by extraneous reasons such as SDN updates, BGP re-routes, and misconfigurations. Such deadlocks can occur despite deadlock-free routing  and must be solved separately (e.g., via sound SDN updates).  

\figput{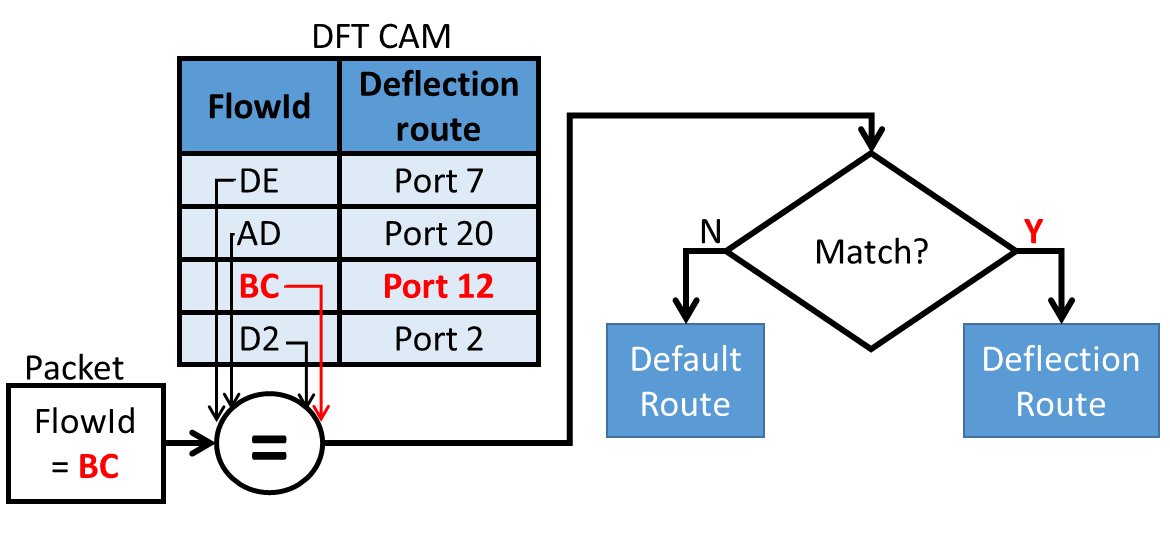}{}{Packet routing with DFT lookup (in hardware)}

IOFD does not misroute long flows. 
Misrouting is a latency optimization for short flows only. Unlike short flows, long flows are sensitive to throughput, not latency. Also, long flows are a dominant fraction of network load, and, therefore, deflecting long flows to longer paths would overload the network.
We achieve this restriction by 
setting  the number of deflection tokens  to zero for long-flow packets.
Deflecting only short flows only a few times
ensures that the increase in link utilization  and path dilation due to IOFD are modest, as  shown in \secref{main}.

Finally, if  IOFD succeeds in dissipating localized
  congestion then DCQCN does not 
kick in (i.e., no ECN marks). Otherwise (e.g., deflection tokens
exhausted), the flows
incur ECN marks which trigger the DCQCN fall-back.
To ensure that IOFD is activated before ECN marks are triggered, 
IOFD's deflection threshold is lower than the ECN threshold.
Our results in \secref{main} show that \name (DASR and IOFD) cuts the number of ECN marks, which trigger  DCQCN fall-backs,
 by 4x  (i.e., the fall-back is infrequent; otherwise, \name would not perform better than DCQCN).

\putsubsec{faulttolerance}{Failures under IOFD}
Because each deflected flow's meta-state is distributed across multiple routers' DFTs, router failures must be correctly handled. To understand how IOFD handles router failures, let us consider how  conventional  RDMA handles failures. 
The back pressure of InfiniBand/RDMA networks ensures that packets queue up at upstream routers (and do not get dropped). 
The neighboring routers detect a failed router and propagate that information back to senders and effectively cause the in-flight packets to be dropped. 
For reliable (i.e., RC) communication, the senders must re-transmit the messages whose completion events have not been received). 
This approach carries over to IOFD without changes irrespective of whether flows have been deflected. As in the baseline case, flows blocked by failures are not allowed to locally reroute around the failed routers (which could cause ordering violations). Instead, all such blocked flows are effectively dropped and must be re-transmitted by the senders.

\putsec{testbed}{Small-scale measurements}
\name has two key components: DASR  which does {\em not} need any hardware switch changes and IOFD  which does. 
Accordingly, 
we implement DASR in our small testbed as we lack access to datacenter-scale networks (this section).
Because hardware changes are hard to implement for a paper, we simulate IOFD, and the full \name, at datacenter scales using ns-3 (\secref{simulation}).

Our testbed consists of 
20 nodes, each consisting of four eight-core AMD Opteron 6320 CPUs running at 2.8 GHz and
256 GB of memory, which connect to a 36-port 
\textit{Mellanox SX6025 InfiniBand} switch using Mellanox ConnectX-3 Pro HCA. 
The  switch provides bidirectional bandwidth of 56 Gbps per port.
All the nodes run RHEL6.7 (kernel version 2.6.32) and  Mellanox OFED 3.3-1.0.4. 

We conduct one experiment to evaluate DASR's convergence (\secref{real-convergence}) and another to evaluate
DASR's performance in the presence of incast (\secref{real-incast}). 
Implementing DASR in our testbed via firmware changes on proprietary NICs is infeasible without vendor support. Fortunately, because we do not have multiple applications in our testbed, we implement AUSS in the application layer, in which
senders and receivers exchange $n$ values using application-level acknowledgements. Further, we evaluate accelerated DASR only
using simulations, and not in our testbed.

\putsubsec{real-convergence}{DASR's convergence}

\figput{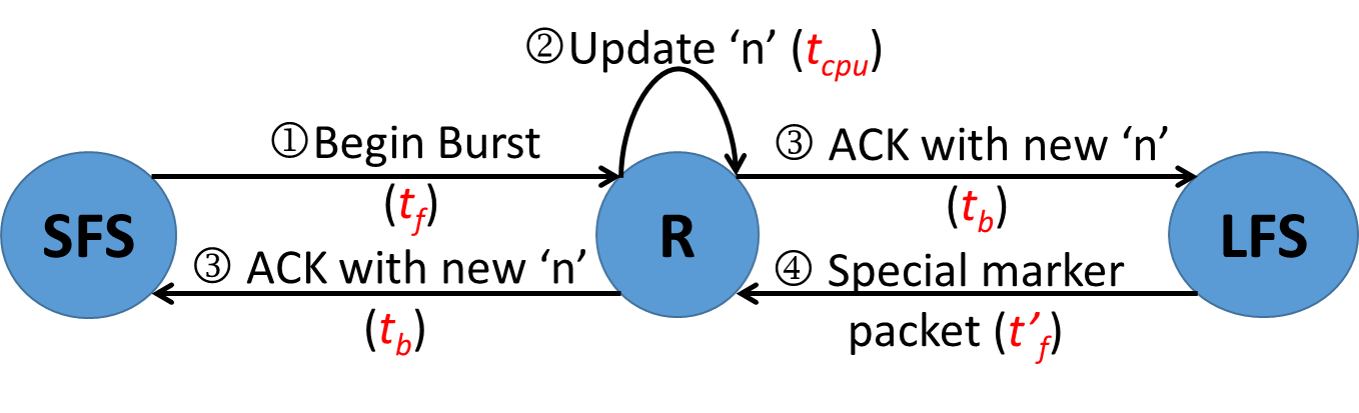}{}{DASR convergence time measurement}

We   answer two key questions: (1) whether DASR converges to fair share bandwidth,
and (2) whether it converges fast. 
We use two senders (\figref{measurement})   -- a long-flow sender (LFS) and a short-flow sender (SFS) --  and a receiver (R). While LFS continuously sends to R, SFS starts a new transmission, taking  $t_f$ to reach R, 
which then takes  $t_{cpu}$ to recalculate the
new $n$ value. Finally, the updated $n$ value is received at  both SFS and LFS,  which then adjust their sending rates, all of which takes  $t_b$.  
The convergence time is the sum of $t_f$, $t_{cpu}$, and $t_b$. However, because the key events occur at different servers with independent clocks, the time components cannot be determined accurately from the events.  
Therefore, we map the  multi-server events into meaningful single-server measurements at R. First, instead of measuring $t_f$, we measure $t_f'$ for a specially-marked message from LFS to R indicating that LFS has seen the new $n$ value.  $t_f$ and $t_f'$ are equal because SFS and LFS are equidistant from R and those paths are not congested (if anything, LFS to R may be loaded more than R to SFS so that $t_f' > t_f$ making our measurements conservative). Second, upon receiving SFS's first message at R, we measure $t_{cpu}$, $t_b$, and $t_f'$, which also add up to the convergence time. 
 
LFS constantly sends 64-KB messages to R. Later, SFS sends 
periodic bursts, during which both SFS and LFS drop to 50\% of the line rate. Each burst consists of 32K messages of 64 KB each. We define the time to send such a burst as an {\em epoch}. 
We measure throughput for groups of 1K messages because per-message bandwidth measurement is extremely noisy.
SFS, LFS and R run on separate nodes.

\figref{real_bw}(a) plots LFS's throughput (Y-axis) over  time in epochs on the X-axis. The vertical grid lines correspond to SFS's bursts. In the absence
of contention, LFS achieves 43 Gbps  which is the peak throughput achieved in our testbed for our message/batch size. However, 
when SFS sends its periodic traffic, LFS
near-instantaneously throttles itself to approximately 
half the sending rate (22.5 Gbps). 
As soon as SFS stops, LFS goes back to the maximum rate. We measured 1K bursts from SFS  (which are seen as troughs in LFS's throughput) but show only five to avoid clutter. 

\figref{real_bw}(a) is not a good indicator of the absolute 
convergence time because the throughput is  averaged over groups of 1K messages. As such, we directly measure DASR's absolute  convergence times in each of the 1024 epochs.
 \figref{real_bw}(b) shows the distribution (solid line) of our 1K measurements of the convergence times (in $\mu$s on X-axis). The $90^{th}$, $99^{th}$, and $99.9^{th}$ percentile convergence times
are 28 $\mu$s, 41 $\mu$s, and
44 $\mu$s, respectively. The unloaded RTT is $15\:\mu s$.
In contrast to DASR's one-RTT convergence,  
TIMELY's convergence takes  50 1-$ms$ RTTs. Figures 18 and 2 in TIMELY~\cite{timely} show  50-$ms$ convergence  and the worst-case RTT to be 1 $ms$, respectively. 
Similarly, RCP~\cite{rcp} and ExpressPass~\cite{expresspass} require several RTTs to 
converge; we study their convergence in \secref{rcp}.

\figput{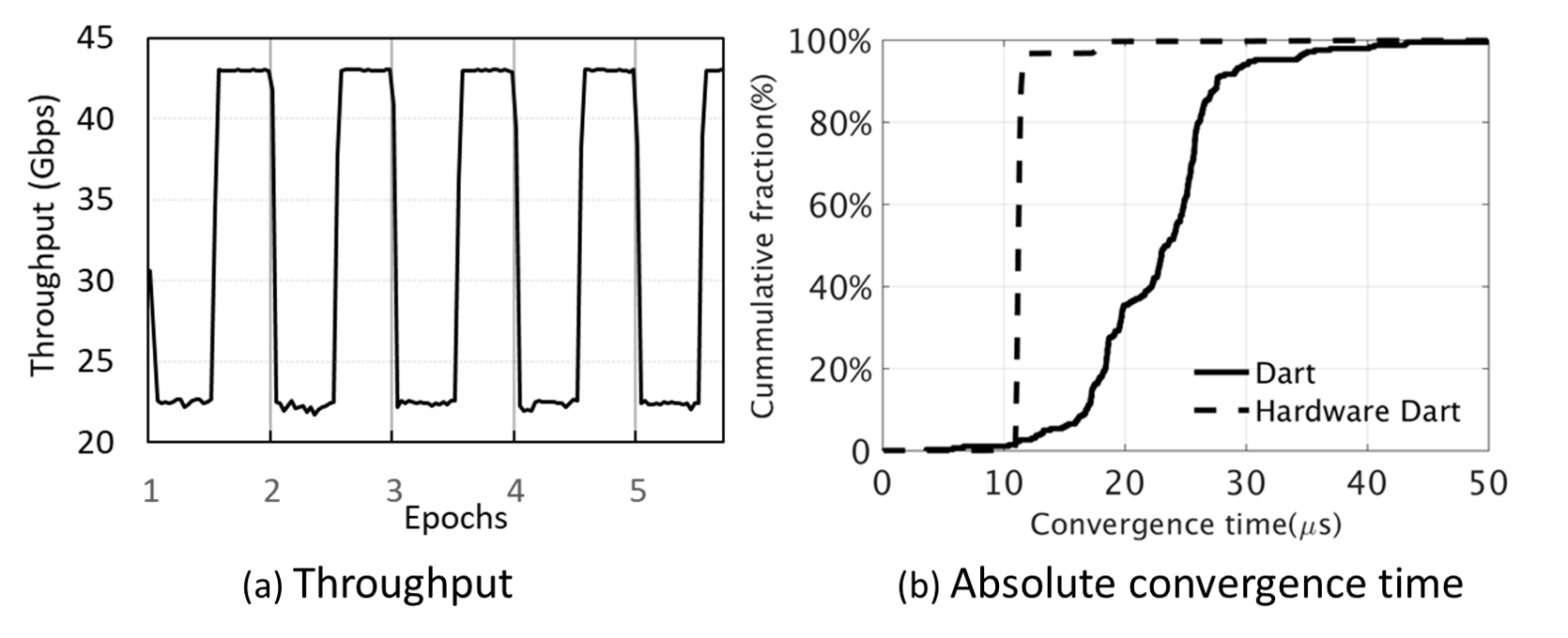}{}{Testbed measurements of DASR convergence}

The above convergence time is for our DASR implementation which maintains the 
AFS in software (\secref{shortflows}). We also show a dashed line in \figref{real_bw}(b) which depicts 
the convergence time for a NIC hardware implementation of DASR. Here,  RDMA's built-in completion queues  notify a sender
that communication is complete which is faster than in software. Then, our convergence time would approach the hardware-RTT (12 $\mu s$). 

\putsubsec{real-incast}{DASR's incast performance}
We compare the completion times of short, incast flows  and throughput
 of long, background flows of InfiniBand and DASR. 
We initiate short 256-KB incasts from a group 
of servers every 100 ms to an {\em aggregator} server. 
Meanwhile, we send continuous background traffic from another server to 
the aggregator. 
We introduce random jitter of 0-100 $\mu$s among  the 
incast senders in each round.
While  InfiniBand uses its congestion control~\cite{mellanox-whitepaper},
we implement DASR's rate control by staggering the messages in time
at the application layer. Here, we do not compare to DCQCN or TIMELY which require NIC firmware changes
and special timer hardware, respectively; we simulate them in~\secref{simulation}. 

\figref{real_fct_cdf} shows the median and tail (99$^{th}$ percentile) flow completion times of DASR and InfiniBand (Y-axis), for varying  incast degrees (X-axis). As expected, 
 higher incast  degrees lead to longer flow completion times  and even longer tails. 
DASR reduces the medians and tails by 2.5 - 3.3x.
DASR's reductions in the tails are close to those in the medians 
because the tails are  only about 1.2x longer than the medians in 
InfiniBand due to our testbed's (small) scale. 
As the tails grow  
at datacenter scales (e.g., 5-10x of the median), DASR  achieves  greater tail reductions (e.g., 5x in \secref{main}).  \figref{real_fct_cdf}(b) shows  the flow completion time 
distributions of InfiniBand and DASR  for the incast degree of 16. 
As compared to InfiniBand,  DASR reduces the spread 
and shifts the curve to the left. 
Both DASR and InfiniBand achieve similar 
throughput (within 0.5\%) for long flows (not shown). 

\figput{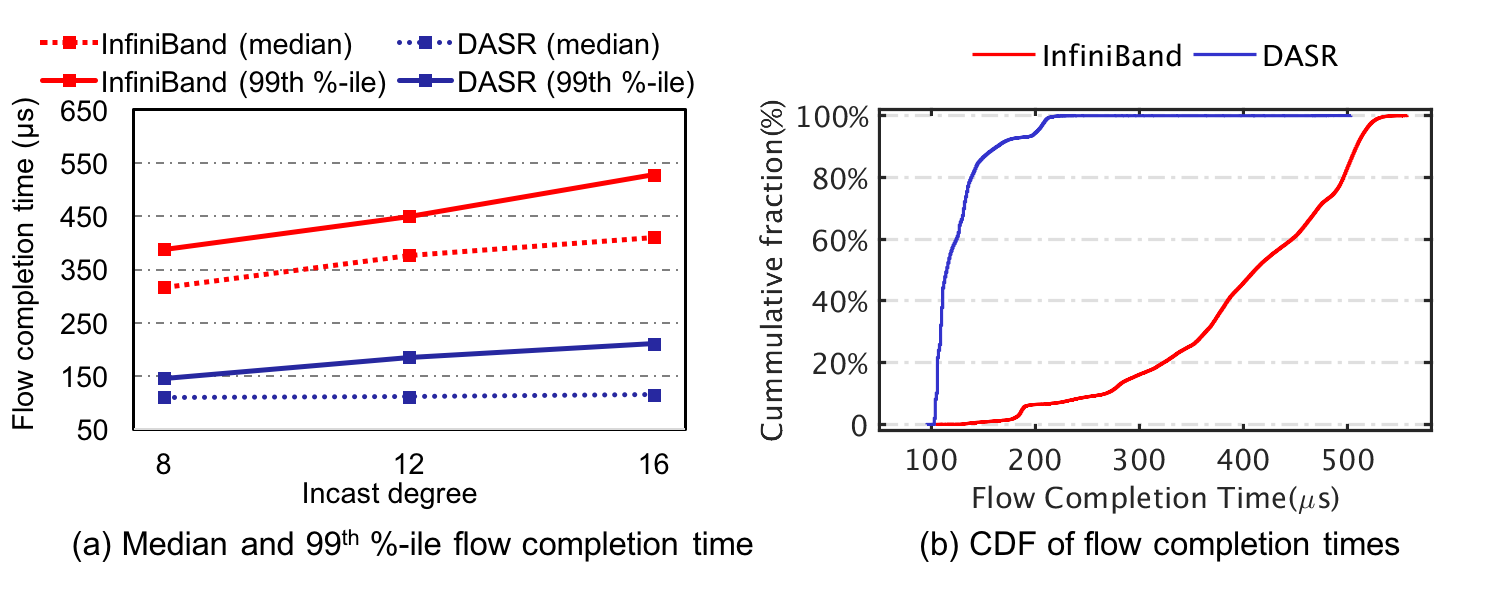}{}{Testbed  flow completion latency}
\putsec{simres}{At-scale simulations}

We evaluate \textit{\name}, DCTCP (includes OS overheads), DCQCN, and TIMELY using typical datacenter traffic patterns~\cite{Benson2010NTC}.

\putsubsec{simulation}{Simulation methodology}

\noindent\textbf{Simulated network:}
We simulate a datacenter with 1024 hosts that are connected in an over-subscribed Clos topology~\cite{clos}. As per common practice, we use (1) an  over-subscription factor of 4 ~\cite{fat-tree-amin}, (2) $10\:Gbps$  point-to-point links with a
propagation delay of $5\:{\mu}s$ so that the longest path is  $6$ hops or $30\:{\mu}s$, and (3) shallow, $225\:KB$ switch buffers and accordingly the ECN threshold of
$22.5\:KB$ (i.e., $10\%$ of the buffer size)~\cite{pfabric, tcpbolt}. 
To utilize all the fat tree paths, we enable Equal Cost Multi-Path (ECMP) routing. \name adaptively deflects packets, in addition to  ECMP. 

\noindent\textbf{Workload}: 
We model our workloads based on real datacenter 
production traffic characteristics~\cite{Benson2010NTC} and similar to TIMELY's~\cite{timely}. Section 4 in~\cite{Benson2010NTC} lists MapReduce and Web applications as the applications that create the traffic. Specifically, 
we follow both the flow size distribution as well as the background/foreground traffic mix from ~\cite{Benson2010NTC}. To model background traffic (e.g., Web Index update), each server initiates a long flow 
of size $1\:GB$ with a randomly-chosen receiver.
Our  foreground traffic that models interactive applications uses short flows of 
size uniformly chosen among \{$2\:KB$, $4\:KB$, and $8\:KB$\} with a default incast-degree of 16 (varied later). Further, groups 
of randomly-chosen servers send to randomly-chosen receivers causing 
multiple incasts which are typical (e.g., in Web Search). 
Further, 
we vary both the overall   network load and the split between background (long) and 
foreground (short, incast) flows.

\noindent\textbf{DCTCP}: Our DCTCP implementation is built over TCP New-Reno. 
We set the initial congestion window to be 10 segments and the
 re-transmit  timeout to $10\:ms$ (typical). We model an 
OS overhead of $300\:{\mu}s$ for each data transfer and calibrate our DCTCP  latencies to match those reported by DCQCN. 

\noindent\textbf{TIMELY}: We implemented TIMELY on {\em ns-3} where the RTT 
measurements are precise (i.e., we avoid the measurement issues
discussed in the TIMELY paper).
While TIMELY uses  64-KB segments to amortize the cost of NIC offload which 
is  not modeled in {\em ns-3}, we use smaller 1460-byte 
segments which provides finer rate control  
and only improves TIMELY's performance in our runs.
To reduce implementation complexity, we use a 
window-based implementation which sets the window size 
based on TIMELY's desired sending rate.
We set TIMELY's parameters as per the TIMELY paper: 
$T_{low}=50\:{\mu}s$, $T_{high}=500\:{\mu}s$, $\alpha=1\:Mbps$, and $\beta=0.8$. We also modeled Hyperactive Increment 
(HAI) for flows to quickly ramp-up their rates.

\figputW{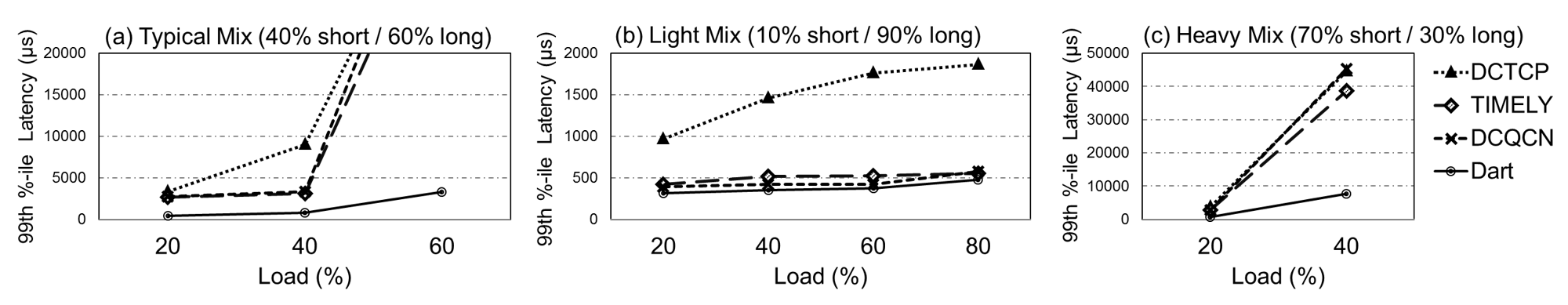}{}{99$^{th}$ percentile flow completion latency}
\figputW{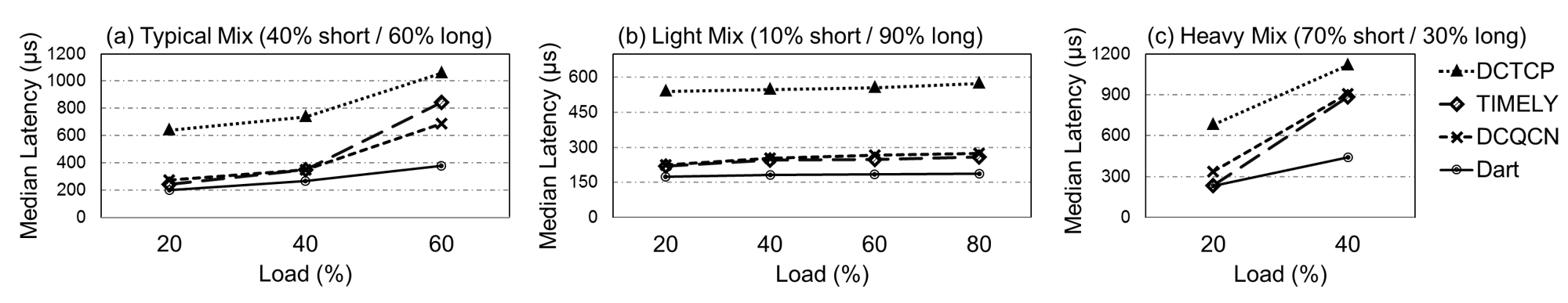}{}{Median flow completion latency}

\noindent\textbf{DCQCN}: 
DCQCN utilizes ECN to infer congestion, similar to DCTCP but with different thresholds. 
On receiving ECN, our simulated receivers run 
the Notification Point (NP) algorithm and 
generate Congestion Notification Packets (CNP)
back to  the sender if needed using high-priority queues. 
The receivers generate at most one CNP 
packet every $50{\mu}s$, as specified by DCQCN.
On receiving a CNP packet, the senders calculate 
their target rate based on DCQCN's Reaction Point (RP)
algorithm.
Following DCQCN's recommendations, we set the exponential averaging factor, 
$g$, to $1/256$, 
the \textit{byte counter} and 
\textit{Timer} to be $10\:MB$ and $55\:{\mu}s$, respectively. 
Flows start at the line rate 
(i.e., there is no slow start). 
Finally, similar to TIMELY's HAI, 
there is a hyper-increase phase to quickly ramp-up the sending rates.

\noindent\textbf{\name}: \name leverages 
DASR and starts  flows at the full line rate (\secref{throughput}). 
We use an 8-entry deflected flow table (DFT);
because we enable IOFD only for short flows (i.e., 2 -- 8 KB flows), 
only a few misrouted flows co-exist at a switch (\secref{mechanisms}).
To ensure that the light-weight IOFD occurs before 
the ECN-based heavy-weight response (\secref{policies}), 
we set the  deflection threshold to be 15 KB
(ECN threshold is 22.5 KB).  Because we experimentally
found that our IOFD's benefits diminish after four misroutes, 
we set the deflection token count to be 4  (\secref{mechanisms}). 

To avoid congestion in the reverse (i.e., \ack)
path for ECN marks in DCTCP and  DCQCN, RTT measurements in TIMELY, and $n$ values in \name, we use high-priority queues only for \ackplural, as suggested by TIMELY. 

\figputW{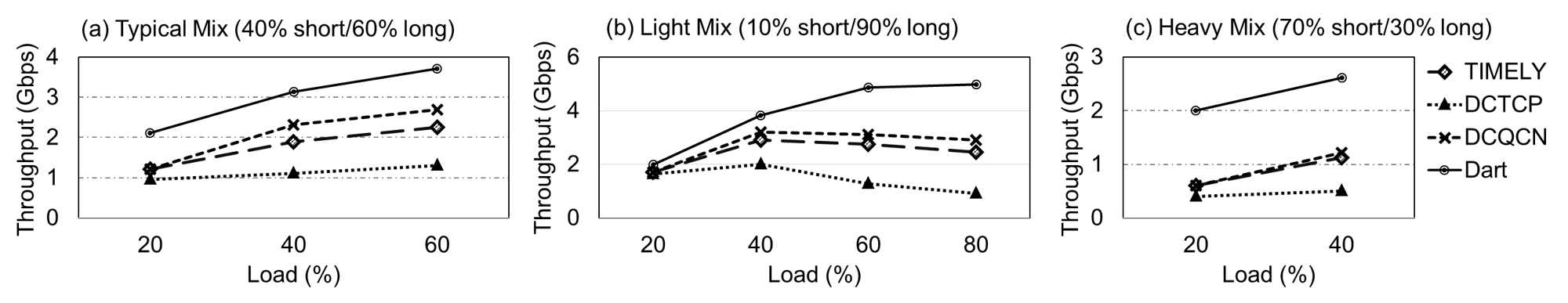}{}{Throughput}

\putsubsec{main}{Latency and throughput}
\figref{99thLatency} plots the 99$^{th}$ percentile flow completion latency (Y-axis) for  all the schemes (individual curves) under  
various load mixes using 8-KB short flows  (the three sub 
graphs) and load levels (X-axis). We show the 8-KB flows out of the
mix of 2-, 4-. and 8-KB flows as described in~\secref{simulation}; we cover the others
in~\secref{isolate}.
Note that the scales of both axes are different for the subgraphs because
the network saturates differently  across load levels.
\figref{50thLatency} is similar to \figref{99thLatency} but it shows
the median latency on the Y-axis.

\textbf{Latency}:
For the typical load-mix (40\% short flows, 60\% long flows), as shown in \figref{99thLatency}(a), 
\name consistently achieves the lowest tail latency at the pre-saturation loads of 20\% and 40\% with a mean reduction in tail latency
of $82\%$ (5.6x); the range 
varies from $79\%$ -- $89\%$ reduction over all the other schemes. 
\name's  (mean) reduction in tail latency is $79\%$ (4.8x) when compared with DCQCN and 
TIMELY (i.e., ignoring DCTCP).  
\name's DASR  avoids iterative convergence for receiver congestion to arrive  accurately   and quickly -- in one RTT -- 
at the appropriate sending rate. {\em We found that 72\% of ECN marks in DCTCP occur at the ToR-receiver links confirming the key result that receiver congestion is the common case ~\cite{KandulaIMC2009,Benson2010NTC,RoySIGCOMM2015,eyeq,QiaoIMC,google-sigcomm15,NDP}.}
Further, \name's IOFD provides quick  response to avoid  spatially-localized in-network congestion. 
Thus, \name's divide-and-specialize approach using 
these two techniques achieves lower latency than TIMELY and DCQCN.
Further,  \name delays the point of saturation past 60\%
load where DCQCN, TIMELY, and DCTCP saturate. DCQCN and TIMELY are similar
because both rely on  iterative convergence  of the 
sending rates  differing only in the congestion signals  
(ECN marks versus RTT measurements as  mentioned in 
\secref{challenges}); their median latencies and throughputs differ more (analyzed later). As expected, both are better than DCTCP, which incurs  high operating system (OS) overhead avoided by the other schemes.

\figref{99thLatency} (b) and (c) illustrate the behavior
when the load mix is made lighter or heavier, respectively,
in terms of short flows (incasts).
For the light load mix (\figref{99thLatency}(b)), 
DCQCN, TIMELY, and \name perform comparably because there is not much room for 
improvement. Due to its high OS overhead, DCTCP's latency penalty remains qualitatively similar to that for the typical load mix.
For the heavy load mix (\figref{99thLatency}(c)),
\name achieves $77\%$ to $83\%$ lower tail latency than the previous schemes.
Further, while 
the previous schemes  saturate above 20\% load, \name's latency increase is 
more modest as \name extends the point of saturation.

One trend  across the load mixes is that 
%the network saturates more readily at higher loads 
the network saturates earlier at higher short-flow fractions. 
This trend is not surprising as short flows do not
offer sufficient time to take reactive action. (On the other hand, 
proactive methods such as slow-start would introduce unnecessary 
latency for short flows.)

\name achieves significantly lower median latency at all load levels and 
load mixes as well (\figref{50thLatency}). On average, \name achieves $30\%$ (1.4x) lower latency than DCQCN and TIMELY and 
$66\%$ (3x) lower latency than DCTCP for the typical load mix.
For the light mix and the heavy  mix (\figref{50thLatency}(b) and \figref{50thLatency}(c), respectively), the latency reductions are $36\%$ 
and $29\%$, respectively.
\name's improvements in median and tail latencies are higher here than in our testbed experiments (\secref{real-incast})
primarily because  the larger scale provides more opportunity.
DCQCN and TIMELY differ modestly in the median latencies in some cases.  
Median latency reduction indicates throughput improvements, as we see next. 

\textbf{Throughput}:
\figref{throughput} shows the throughput achieved for the same set of load levels and load-mix ratios.
\figref{throughput}(a) shows that \name consistently outperforms the DCQCN and TIMELY. The mean improvement in throughput is $48\%$ and $68\%$, 
(mean across all load levels) over DCQCN and TIMELY, respectively.  DASR's accurate and one-RTT convergence is the key reason for \name's  higher throughput. IOFD directly improves only the latency and   affects the throughput only indirectly by avoiding  DCQCN fall-back which would cut the sending rates.
 As with latency, DCQCN and TIMELY outperform DCTCP in throughput due to DCTCP's OS overheads. 
With the heavy mix (\figref{throughput}(c)), \name is $173\%$ better (on average) than DCQCN and TIMELY. 
This improvement is not surprising as both
DCQCN and TIMELY saturate at such heavy loads.
Though the relative ordering  with the light mix (\figref{throughput}(b)) remains the same as that with the typical mix, the 
absolute throughputs are higher, as expected. We  see the correspondence between \name's median latency and throughput at high loads. Like the median latencies of DCQCN and TIMELY, their throughputs also differ slightly.

\tabput{table-ecn}{
\small
\begin{tabular}{|c|c|c|c|c|c|c|c|c|c|}
\hline
\multirow{2}{*}{\begin{tabular}[c]{@{}c@{}}Traffic mix\\ / Load\end{tabular}} & \multicolumn{3}{c|}{Typical Mix} & \multicolumn{4}{c|}{Light Mix}   & \multicolumn{2}{c|}{Heavy Mix} \\ \cline{2-10}                                                         

& 20 & 40 & 60 & 
  20 & 40 & 60 & 80 & 
  20 & 40
\\ \hline

DCQCN & 17 & 41 & 67 & 
  5 & 14 & 36 & 48 & 
  33 & 70
\\ \hline

\name & 6 & 11 & 14 & 
  4 & 9 & 21 & 28 & 
  9 & 18
\\ \hline
\end{tabular}
}{\% short-flow packets with ECN marks}

\noindent
\textbf{Fall-back to DCQCN}:
To evaluate  DCQCN fall-backs in \name, \tabref{table-ecn} shows the percent of short-flow packets with ECN marks under
DCQCN and \name. Because long flows are not latency-critical, we focus on short flows.  As expected,
both schemes incur more ECN marks as the load increases. However, \name cuts the number of ECN marks by more than 4x at higher loads in typical and heavy mixes (i.e., significant fraction of short flows) where there is more congestion. These  results (1) show that by filtering out 
receiver congestion and localized in-network congestion, \name drastically reduces the number of DCQCN fall-backs and (2) reconfirm that these congestion components are
significant.

\noindent 
\textbf{Load increase and path dilation due to IOFD}:
\tabref{table-pathdilation} shows the percent increase in (a) network load and (b) short-flow path length under IOFD relative to DCQCN. Both the load
and path dilation increase with more short flows ((i.e., light < typical < heavy) 
and at higher loads. For typical and heavy mixes, \name increases the network load by 7\%  (geometric mean over the load settings) and dilates short-flow paths by 16\% which is roughly one hop (our topology has 5.8 hops on average).
Thus, \name incurs a modest amount of network load to reduce congestion delays significantly.

\tabput{table-pathdilation}{
\small
\begin{tabular}{|c|c|c|c|c|c|c|c|c|c|}
\hline
\multirow{2}{*}{\begin{tabular}[c]{@{}c@{}}Traffic mix\\ / Load\end{tabular}} & \multicolumn{3}{c|}{Typical } & \multicolumn{4}{c|}{Light }   & \multicolumn{2}{c|}{Heavy } \\ \cline{2-10} 
                                                                     & 20 & 40 & 60 & 
  20 & 40 & 60 & 80 & 
  20 & 40
\\ \hline

Load                                                          & \multirow{2}{*}{3} & \multirow{2}{*}{6} & \multirow{2}{*}{9} & 
  \multirow{2}{*}{0.4}& \multirow{2}{*}{0.8} & \multirow{2}{*}{1.4} & \multirow{2}{*}{3} & 
  \multirow{2}{*}{11} & \multirow{2}{*}{15}        
\\ 
increase & & 
& &
& & & &
& \\
\hline
Path                    
& \multirow{2}{*}{8} & \multirow{2}{*}{15} & \multirow{2}{*}{23} & 
  \multirow{2}{*}{4} & \multirow{2}{*}{8} & \multirow{2}{*}{14} & \multirow{2}{*}{26} & 
  \multirow{2}{*}{15} & \multirow{2}{*}{21}
\\
dilation & & & & & & & & &\\
\hline
\end{tabular}
}{IOFD's load increase and path dilation (\%)}

\putsubsec{isolate}{Isolating \name's techniques}

We quantify the relative contributions of \name's two techniques: DASR and IOFD.
\figref{isolate} plots \name's $99^{th}$ percentile flow completion latency for the 8-KB short flows normalized to that of DCQCN (Y-axis)
for the typical load mix  (i.e., 40\% short flows and 60\% long flows) at various load levels (groups of bars along the X-axis).
In addition to \name, we quantify the benefits of 
DCQCN with priority queues using two priority levels to prioritize short flows over long flows ({\em Pri-Q}),
IOFD without DASR ({\em IOFD-only}), 
DASR without IOFD or the look-ahead optimization  in \secref{lookahead} ({\em DASR w/o LA}),  and DASR with look-ahead but without IOFD ({\em DASR-only}).

As we see from \figref{isolate}, 
Pri-Q does not improve latency at low loads where the long flows do not cause much congestion and hence provide limited opportunity. At higher loads, Pri-Q improves latency as expected. 
However, the improvement is limited because Pri-Q does not alleviate congestion  \textit{among} short flows, 
and, therefore,  performs worse than IOFD-only and DASR-only, our key techniques. 
IOFD and DASR specifically address congestion among short flows --- IOFD addresses localized congestion, whereas DASR addresses receiver congestion. 
Three key trends regarding the relative benefits of DASR and IOFD are apparent in \figref{isolate}.
First, at the intermediate load of 40\% (middle bars), each of DASR (including the look-ahead) and IOFD contribute to \name's  improvements. DASR contributes more because receiver congestion is the common case (\secref{main}). Further, the  difference between DCQCN and  {\em IOFD-only} shows that IOFD  can handle localized congestion without triggering  DCQCN fall-back (\secref{policies}). 
Second, at lower loads (left bars), most of the gains  come from DASR which effectively protects  the short flows from
the long flows ($79\%$ of the $84\%$ total latency reduction). This result is not surprising because in-network congestion is less likely at lower loads.  The sizable difference between {\em DASR w/o LA} and {\em DASR-only} shows the look-ahead's impact. 
In the absence of localized  congestion, the opportunity for IOFD is lower; as such {\em IOFD-only}  contributes 
only $31\%$ latency reduction in isolation, and approximately $21\%$ incremental latency reduction over {\em DASR-only}.
Finally, in contrast to the low-load results, 
IOFD contributes  relatively more to the overall latency reduction  at higher loads, 
 where in-network congestion is more likely ($79\%$ of the $93\%$ total 
latency reduction at 60\% load). IOFD handles even this  higher congestion without falling back to  DCQCN.  {\em DASR-only}'s relative contribution  is smaller  than
 IOFD's ($77\%$ latency reduction in isolation, and $35\%$ incremental latency over {\em IOFD-only}). The median latencies follow the same trends. The effectiveness of {\em DASR-only} and {\em IOFD-only} illustrate the power of \name's divide-and-specialize approach. 
 
\figput{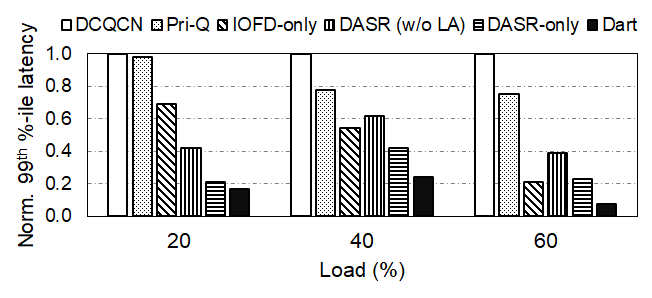}{}{Isolating \name's techniques}

\noindent\textbf{Sensitivity}: 
We varied the deflection threshold (\secref{policies}) as  5, 15 (default) and 20 KB. IOFD works well in the range of 5-15 KB, whereas the 20-KB threshold being close to the ECN threshold (22.5 KB) results in  IOFD  being disabled. We also varied the short-flow sizes as 2, 4 and 8 (default) KB. \name's  
improvement across these flow sizes match those in \figref{99thLatency}. Finally, we varied the incast degree as $6$, $16$ 
(default), and $26$. At higher incast  degree, \name's latency improvement over DCQCN increases. However, at 60\% load and incast degree of $26$, the network saturates leaving no room for \name.  These results are not shown due to lack of space.

\putsubsec{rcp}{Comparison to RCP and ExpressPass}

\figput[1.0\linewidth]{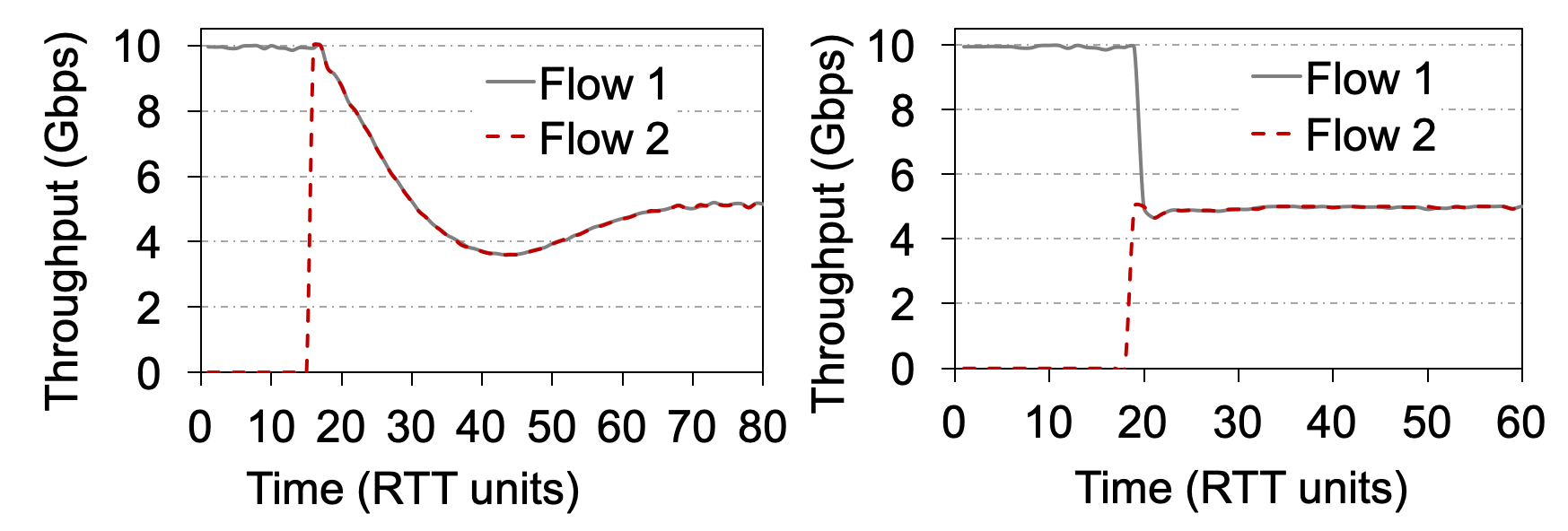}{}{Convergence time for RCP and \name}

\paragraph{\textbf{Comparison to RCP}} We study RCP's convergence using an  ns-3 implementation~\cite{rcp-router}. We simulate a simple topology with three servers that connect to a single switch. The topology uses $10~Gbps$ links with a delay of $5~{\mu}s$, matching today's datacenters. We set RCP's main parameters, $\alpha = 0.4$ and $\beta = 0.4$, as recommended~\cite{rcp, nandita-thesis}. 

We compare how fast RCP and \name  converge to the fair-share rate of $5 Gbps$ for the simple case of receiver congestion with two flows. In our experiment, the second flow starts after the first flow. \figref{rcp} shows time, in RTT time units, along the X axis and the throughputs achieved by the two flows for RCP (in \figref{rcp}(a)) and \name (in \figref{rcp}(b)), along the Y axis. Because RCP does not \emph{explicitly} count flows, it requires about $60~RTTs$ to converge (i.e., the second flow joins at about $15$ RTTs but the rate does not converge to $5~Gbps$ until $75$ RTTs). In contrast, \name converges in $1 RTT$ by \emph{explicitly} tracking the number of senders at the receiver. 

\paragraph{\textbf{Comparison to ExpressPass}} We obtained the ExpressPass~\cite{expresspass} simulator from the authors. We found that while  ExpressPass converges in 4 RTTs for two senders and two receivers in a simple dumbbell topology (matches ExpressPass paper's results), it takes 20 RTTs for  10 senders and one receiver in the fat tree topology used in the paper (the paper does not show this case).
In the former case, convergence is effected by fair queuing of credit packets at the switch where the two flows collide, whereas the latter case converges using ExpressPass's BIC-TCP-like algorithm,
which is iterative and slow. 
\putsec{related}{Related Work}
Because we have discussed  DCQCN, TIMELY,
NUMFabric, ExpressPass, and NDP at length
in earlier sections, we focus on other work
related to our key techniques --  DASR (congestion
control) and IOFD (load balancing).

DCTCP~\cite{dctcp}, a pioneering work in datacenter transport protocols, 
finely modulates the sending rate by observing  ECN marks in each RTT 
and nearly eliminates incast-induced timeouts. 
$D^2$TCP~\cite{d2tcp} builds upon DCTCP  to prioritize flows based on deadlines.
TCP Bolt~\cite{tcpbolt} uses flow-level congestion control via ECN to 
address PFC's limitations. ICTCP~\cite{ICTCP} iteratively adjusts the  TCP receive window before incast-induced packet drops.
Like DCQCN and TIMELY, all these TCP variants incur several RTTs (i.e., tens or hundreds) to converge to the appropriate sending rate.  
In RCP~\cite{rcp}, another pioneering work, routers explicitly convey  the fair share rate to  
the senders that share a  link.
However, because RCP routers don't have per-flow state and 
  many short flows begin and end in each RTT, 
RCP's convergence is iterative and takes many RTTs.
D3~\cite{d3} and PDQ~\cite{pdq} employs explicit rate control  to prioritize critical flows. 
For the same reasons as RCP, D3 and PDQ's convergence requires several RTTs.
pFabric~\cite{pfabric}, Karuna~\cite{karuna}, 
UPS~\cite{ups} and pHost~\cite{phost} address  
flow scheduling but their rate control is still iterative. 
EyeQ~\cite{eyeq} leverages RCP to provide weighted fair share in
multi-tenant datacenters but inherits RCP's iterative convergence. 
 QCN~\cite{qcn} provides end-to-end congestion control for RDMA in Layer2 but not in 
IP-switched  datacenter-scale networks. 
IRN~\cite{mittal-rdma} adapts ideas from I-WARP to avoid the problems associated with PFC.
Unlike DASR, none of the above work isolates receiver congestion to achieve 
fast, \textit{one-RTT} convergence. 
DIATCP~\cite{diatcp} is a receiver-based rate control protocol where each sender informs the receiver of the message size and deadline so that the receiver calculates the per-sender sending rate accordingly. Though DIATCP can achieve one-RTT convergence, its calculation requires an a priori  globally-common RTT which is achieved by artificially delaying {\em acks}. In a DC setting, such delay would force {\em all} flows to have the same RTT as the tail, which is often 2-5x longer than the median~\cite{dctcp}. In contrast, \name sends only the sender count to the senders each of which then flexibly and locally arrives at a sending rate based on the currently-observed, individual RTT. 
Apart from this fundamental difference, while both \name and DIATCP use timeouts (\secref{shortflows}) they do so for different purposes. Because RDMA has explicit message start/end markers, \name does not need to use timeouts for message ends. Instead, \name  uses timeouts for  infrequent, catastrophic events like crashes. In contrast, DIATCP uses timeouts to detect idling because TCP does not have message start/end markers. Thus, \name's mechanisms are optimized  for RDMA unlike  DIATCP.  Because flow idling  is more frequent than crashes, DIATCP's timeouts can affect performance -- a timeout being too short means delayed flow completions and  being too long means wasted throughput. Further, each sender in DIATCP provides its deadline and message size to the receiver at the time of establishing the connection. This exchange occurs before each flow starts, but not necessarily before the start of all the other flows in an incast group.  As such, DIATCP cannot know the correct rate before {\em all} the flows start.
In contrast, DASR's lookahead (\secref{lookahead}) is  based on the application  where all the flows in an incast group know before {\em any} of them  start that they are a part of an incast group. Thus, DIATCP does not achieve DASR's lookahead. 

Among load balancing schemes,  MPTCP~\cite{MPTCPSIGCOMM11,MPTCPNSDI12}
splits a TCP flow into many sub-flows that may be routed 
independently along different paths.
FlowBender~\cite{flowbender} proposes re-hashing at end-hosts to change flow paths.
Presto~\cite{Presto16} splits large flows into equal-sized flowcells 
and uses a central scheduler to balance the load. 
DeTail~\cite{detail}, Random Packet Spraying~\cite{pktscatterinfocom},
and DIBS~\cite{dibs} 
balance load at the finer granularity of packets. 
Recent schemes~\cite{hermes-sigcomm17,drill-sigcomm17} improve load balancing but they 
also require reordering at the receiver.
In contrast to these above schemes all of which reorder packets which is not supported by RDMA,  IOFD is designed to deflect without packet reordering. 
SPAIN~\cite{spain}  and CONGA~\cite{CONGA}  also avoid packet reordering. However, 
SPAIN pre-computes multiple paths which are mapped to different VLANs but such precomputation may be
slow in reaction to short flows in a datacenter. 
CONGA uses global congestion information to 
load balance at the granularity of flowlets. 
However, CONGA  works only with two-tier leaf-spine 
topology and does not scale to large datacenters.

\putsec{concl}{Conclusion}

RDMA can significantly reduce datacenter network latencies compared to TCP but provides suboptimal end-to-end
congestion control for the well-known problem of incasts. Previous schemes target the full generality of the congestion problem and rely on slow, iterative convergence to the appropriate
sending rates. Several papers have shown that  even in oversubscribed datacenter networks most congestion occurs at the receiver. Accordingly, we proposed  a divide-and-specialize approach, called {\em \name}, which isolates the common case of receiver congestion and further sub-divides the remaining in-network congestion into the simpler spatially-localized  and the harder spatially-dispersed cases. To address receiver congestion,  we proposed {\em direct apportioning of sending rates (DASR)} in which  a receiver for $n$ senders directs each sender to cut its rate by a factor of $n$. DASR converges   in only one RTT.
For the spatially-localized case, \name adds novel switch hardware  for {\em in-order flow deflection (IOFD)} because RDMA disallows packet reordering on which previous load balancing schemes rely. IOFD provides fast (under one RTT), light-weight response. For the uncommon spatially-dispersed case, \name falls back to DCQCN. 
Our small-scale testbed measurements showed that \name  converges in one RTT and achieves  $60\%$ (2.5x) lower tail ($99^{th}$-percentile) latency than and similar throughput as InfiniBand. Our at-scale simulations  showed that 
\name achieves  $79\%$ (4.8x) lower tail latency, and 
$58\%$ higher throughput than TIMELY and DCQCN.
As datacenter networks evolve towards adopting RDMA to avoid TCP's overhead,
\name's superior latency and throughput characteristics are likely to be
attractive.

%\section*{Acknowledgments}
\noindent
{\bf Acknowledgments:} This work was supported in part by the National Science Foundation under grant numbers 1633412-IIS and 1633318-IIS.

\bibliographystyle{IEEEtran}

\bibliography{local,references}

% Generated by IEEEtran.bst, version: 1.14 (2015/08/26)
\begin{thebibliography}{10}
\providecommand{\url}[1]{#1}
\csname url@samestyle\endcsname
\providecommand{\newblock}{\relax}
\providecommand{\bibinfo}[2]{#2}
\providecommand{\BIBentrySTDinterwordspacing}{\spaceskip=0pt\relax}
\providecommand{\BIBentryALTinterwordstretchfactor}{4}
\providecommand{\BIBentryALTinterwordspacing}{\spaceskip=\fontdimen2\font plus
\BIBentryALTinterwordstretchfactor\fontdimen3\font minus
  \fontdimen4\font\relax}
\providecommand{\BIBforeignlanguage}[2]{{%
\expandafter\ifx\csname l@#1\endcsname\relax
\typeout{** WARNING: IEEEtran.bst: No hyphenation pattern has been}%
\typeout{** loaded for the language `#1'. Using the pattern for}%
\typeout{** the default language instead.}%
\else
\language=\csname l@#1\endcsname
\fi
#2}}
\providecommand{\BIBdecl}{\relax}
\BIBdecl

\bibitem{MPI-infiniband}
\BIBentryALTinterwordspacing
J.~Liu, J.~Wu, and D.~K. Panda, ``High performance rdma-based mpi
  implementation over infiniband,'' \emph{Int. J. Parallel Program.}, vol.~32,
  no.~3, pp. 167--198, Jun. 2004. [Online]. Available:
  \url{http://dx.doi.org/10.1023/B:IJPP.0000029272.69895.c1}
\BIBentrySTDinterwordspacing

\bibitem{MICA}
\BIBentryALTinterwordspacing
H.~Lim, D.~Han, D.~G. Andersen, and M.~Kaminsky, ``Mica: A holistic approach to
  fast in-memory key-value storage,'' in \emph{Proceedings of the 11th USENIX
  Conference on Networked Systems Design and Implementation}, ser.
  NSDI'14.\hskip 1em plus 0.5em minus 0.4em\relax Berkeley, CA, USA: USENIX
  Association, 2014, pp. 429--444. [Online]. Available:
  \url{http://dl.acm.org/citation.cfm?id=2616448.2616488}
\BIBentrySTDinterwordspacing

\bibitem{pilaf}
\BIBentryALTinterwordspacing
C.~Mitchell, Y.~Geng, and J.~Li, ``Using one-sided rdma reads to build a fast,
  cpu-efficient key-value store,'' in \emph{Proceedings of the 2013 USENIX
  Conference on Annual Technical Conference}, ser. USENIX ATC'13.\hskip 1em
  plus 0.5em minus 0.4em\relax Berkeley, CA, USA: USENIX Association, 2013, pp.
  103--114. [Online]. Available:
  \url{http://dl.acm.org/citation.cfm?id=2535461.2535475}
\BIBentrySTDinterwordspacing

\bibitem{FaRM}
\BIBentryALTinterwordspacing
A.~Dragojevi{\'c}, D.~Narayanan, O.~Hodson, and M.~Castro, ``Farm: Fast remote
  memory,'' in \emph{11th USENIX Symposium on Networked Systems Design and
  Implementation (NSDI 2014)}.\hskip 1em plus 0.5em minus 0.4em\relax USENIX
  – Advanced Computing Systems Association, April 2014. [Online]. Available:
  \url{http://research.microsoft.com/apps/pubs/default.aspx?id=208395}
\BIBentrySTDinterwordspacing

\bibitem{HERD}
\BIBentryALTinterwordspacing
A.~Kalia, M.~Kaminsky, and D.~G. Andersen, ``Using rdma efficiently for
  key-value services,'' in \emph{Proceedings of the 2014 ACM Conference on
  SIGCOMM}, ser. SIGCOMM '14.\hskip 1em plus 0.5em minus 0.4em\relax New York,
  NY, USA: ACM, 2014, pp. 295--306. [Online]. Available:
  \url{http://doi.acm.org/10.1145/2619239.2626299}
\BIBentrySTDinterwordspacing

\bibitem{Panda-memcached}
\BIBentryALTinterwordspacing
J.~Jose, H.~Subramoni, K.~Kandalla, M.~Wasi-ur Rahman, H.~Wang, S.~Narravula,
  and D.~K. Panda, ``Scalable memcached design for infiniband clusters using
  hybrid transports,'' in \emph{Proceedings of the 2012 12th IEEE/ACM
  International Symposium on Cluster, Cloud and Grid Computing (Ccgrid 2012)},
  ser. CCGRID '12.\hskip 1em plus 0.5em minus 0.4em\relax Washington, DC, USA:
  IEEE Computer Society, 2012, pp. 236--243. [Online]. Available:
  \url{http://dx.doi.org/10.1109/CCGrid.2012.141}
\BIBentrySTDinterwordspacing

\bibitem{RDMA-IPDPS10}
E.~Gran, M.~Eimot, S.-A. Reinemo, T.~Skeie, O.~Lysne, L.~Huse, and G.~Shainer,
  ``First experiences with congestion control in infiniband hardware,'' in
  \emph{Parallel Distributed Processing (IPDPS), 2010 IEEE International
  Symposium on}, April 2010, pp. 1--12.

\bibitem{mellanox-whitepaper}
D.~Crupnicoff, S.~Das, and E.~Zahavi, ``{White paper: Deploying Quality of
  Service and Congestion Control in InfiniBand-based Data Center Networks},''
  Mellanox Technologies, Tech. Rep. 2379, November 2005.

\bibitem{dctcp}
\BIBentryALTinterwordspacing
M.~Alizadeh, A.~Greenberg, D.~A. Maltz, J.~Padhye, P.~Patel, B.~Prabhakar,
  S.~Sengupta, and M.~Sridharan, ``Data center tcp (dctcp),'' in
  \emph{Proceedings of the ACM SIGCOMM 2010 conference}, ser. SIGCOMM
  '10.\hskip 1em plus 0.5em minus 0.4em\relax New York, NY, USA: ACM, 2010, pp.
  63--74. [Online]. Available: \url{http://doi.acm.org/10.1145/1851182.1851192}
\BIBentrySTDinterwordspacing

\bibitem{fat-tree-amin}
\BIBentryALTinterwordspacing
M.~Al-Fares, A.~Loukissas, and A.~Vahdat, ``A scalable, commodity data center
  network architecture,'' in \emph{Proceedings of the ACM SIGCOMM 2008
  conference on Data communication}, ser. SIGCOMM '08.\hskip 1em plus 0.5em
  minus 0.4em\relax New York, NY, USA: ACM, 2008, pp. 63--74. [Online].
  Available: \url{http://doi.acm.org/10.1145/1402958.1402967}
\BIBentrySTDinterwordspacing

\bibitem{fat-tree-original}
\BIBentryALTinterwordspacing
C.~E. Leiserson, ``Fat-trees: universal networks for hardware-efficient
  supercomputing,'' \emph{IEEE Trans. Comput.}, vol.~34, no.~10, pp. 892--901,
  Oct. 1985. [Online]. Available:
  \url{http://dl.acm.org/citation.cfm?id=4492.4495}
\BIBentrySTDinterwordspacing

\bibitem{dcqcn}
\BIBentryALTinterwordspacing
Y.~Zhu, H.~Eran, D.~Firestone, C.~Guo, M.~Lipshteyn, Y.~Liron, J.~Padhye,
  S.~Raindel, M.~H. Yahia, and M.~Zhang, ``Congestion control for large-scale
  rdma deployments,'' in \emph{Proceedings of the 2015 ACM Conference on
  Special Interest Group on Data Communication}, ser. SIGCOMM '15.\hskip 1em
  plus 0.5em minus 0.4em\relax ACM, 2015, pp. 523--536. [Online]. Available:
  \url{http://doi.acm.org/10.1145/2785956.2787484}
\BIBentrySTDinterwordspacing

\bibitem{timely}
\BIBentryALTinterwordspacing
R.~Mittal, V.~T. Lam, N.~Dukkipati, E.~Blem, H.~Wassel, M.~Ghobadi, A.~Vahdat,
  Y.~Wang, D.~Wetherall, and D.~Zats, ``Timely: Rtt-based congestion control
  for the datacenter,'' in \emph{Proceedings of the 2015 ACM Conference on
  Special Interest Group on Data Communication}, ser. SIGCOMM '15.\hskip 1em
  plus 0.5em minus 0.4em\relax New York, NY, USA: ACM, 2015, pp. 537--550.
  [Online]. Available: \url{http://doi.acm.org/10.1145/2785956.2787510}
\BIBentrySTDinterwordspacing

\bibitem{Benson2010NTC}
\BIBentryALTinterwordspacing
T.~Benson, A.~Akella, and D.~A. Maltz, ``Network traffic characteristics of
  data centers in the wild,'' in \emph{Proceedings of the 10th ACM SIGCOMM
  conference on Internet measurement}, ser. IMC '10.\hskip 1em plus 0.5em minus
  0.4em\relax New York, NY, USA: ACM, 2010, pp. 267--280. [Online]. Available:
  \url{http://doi.acm.org/10.1145/1879141.1879175}
\BIBentrySTDinterwordspacing

\bibitem{eyeq}
V.~Jeyakumar, M.~Alizadeh, D.~Mazi\`{e}res, B.~Prabhakar, C.~Kim, and
  A.~Greenberg, ``Eyeq: Practical network performance isolation at the edge,''
  in \emph{Proceedings of the 10th USENIX Conference on Networked Systems
  Design and Implementation}, ser. nsdi'13, 2013, pp. 297--312.

\bibitem{QiaoIMC}
Q.~Zhang, V.~Liu, H.~Zeng, and A.~Krishnamurthy, ``High-resolution measurement
  of data center microbursts,'' in \emph{Proceedings of the 2017 Internet
  Measurement Conference}, ser. IMC '17.\hskip 1em plus 0.5em minus 0.4em\relax
  ACM, 2017, pp. 78--85.

\bibitem{NDP}
M.~Handley, C.~Raiciu, A.~Agache, A.~Voinescu, A.~W. Moore, G.~Antichi, and
  M.~W\'{o}jcik, ``Re-architecting datacenter networks and stacks for low
  latency and high performance,'' in \emph{Proceedings of the Conference of the
  ACM Special Interest Group on Data Communication}, ser. SIGCOMM '17.\hskip
  1em plus 0.5em minus 0.4em\relax ACM, 2017, pp. 29--42.

\bibitem{RoySIGCOMM2015}
A.~Roy, H.~Zeng, J.~Bagga, G.~Porter, and A.~C. Snoeren, ``Inside the social
  network's (datacenter) network,'' in \emph{Proceedings of the 2015 ACM
  Conference on Special Interest Group on Data Communication}, ser. SIGCOMM
  '15.\hskip 1em plus 0.5em minus 0.4em\relax ACM, 2015, pp. 123--137.

\bibitem{google-sigcomm15}
A.~Singh, J.~Ong, A.~Agarwal, G.~Anderson, A.~Armistead, R.~Bannon, S.~Boving,
  G.~Desai, B.~Felderman, P.~Germano, A.~Kanagala, J.~Provost, J.~Simmons,
  E.~Tanda, J.~Wanderer, U.~H\"{o}lzle, S.~Stuart, and A.~Vahdat, ``Jupiter
  rising: A decade of clos topologies and centralized control in google's
  datacenter network,'' in \emph{Proceedings of the 2015 ACM Conference on
  Special Interest Group on Data Communication}, ser. SIGCOMM '15.\hskip 1em
  plus 0.5em minus 0.4em\relax ACM, 2015, pp. 183--197.

\bibitem{KandulaIMC2009}
S.~Kandula, S.~Sengupta, A.~Greenberg, P.~Patel, and R.~Chaiken, ``The nature
  of data center traffic: Measurements \& analysis,'' in \emph{Proceedings of
  the 9th ACM SIGCOMM Conference on Internet Measurement}, ser. IMC '09.\hskip
  1em plus 0.5em minus 0.4em\relax ACM, 2009, pp. 202--208.

\bibitem{MPTCPSIGCOMM11}
\BIBentryALTinterwordspacing
C.~Raiciu, S.~Barre, C.~Pluntke, A.~Greenhalgh, D.~Wischik, and M.~Handley,
  ``Improving datacenter performance and robustness with multipath tcp,'' in
  \emph{Proceedings of the ACM SIGCOMM 2011 conference}, ser. SIGCOMM
  '11.\hskip 1em plus 0.5em minus 0.4em\relax New York, NY, USA: ACM, 2011, pp.
  266--277. [Online]. Available:
  \url{http://doi.acm.org/10.1145/2018436.2018467}
\BIBentrySTDinterwordspacing

\bibitem{spain}
J.~Mudigonda, P.~Yalagandula, M.~Al-Fares, and J.~C. Mogul, ``Spain: Cots
  data-center ethernet for multipathing over arbitrary topologies,'' in
  \emph{Proceedings of the 7th USENIX Conference on Networked Systems Design
  and Implementation}, ser. NSDI'10, 2010.

\bibitem{pktscatterinfocom}
A.~Dixit, P.~Prakash, Y.~Hu, and R.~Kompella, ``On the impact of packet
  spraying in data center networks,'' in \emph{INFOCOM, 2013 Proceedings IEEE},
  2013, pp. 2130--2138.

\bibitem{detail}
\BIBentryALTinterwordspacing
Z.~et~al., ``Detail: reducing the flow completion time tail in datacenter
  networks,'' in \emph{Proceedings of the ACM SIGCOMM 2012 conference on
  Applications, technologies, architectures, and protocols for computer
  communication}, ser. SIGCOMM '12.\hskip 1em plus 0.5em minus 0.4em\relax New
  York, NY, USA: ACM, 2012, pp. 139--150. [Online]. Available:
  \url{http://doi.acm.org/10.1145/2342356.2342390}
\BIBentrySTDinterwordspacing

\bibitem{flowbender}
A.~Kabbani, B.~Vamanan, J.~Hasan, and F.~Duchene, ``Flowbender: Flow-level
  adaptive routing for improved latency and throughput in datacenter
  networks,'' in \emph{Proceedings of the 10th ACM International on Conference
  on Emerging Networking Experiments and Technologies}, ser. CoNEXT '14, 2014,
  pp. 149--160.

\bibitem{Presto16}
K.~He, E.~Rozner, K.~Agarwal, W.~Felter, J.~Carter, and A.~Akella, ``Presto:
  Edge-based load balancing for fast datacenter networks,'' in
  \emph{Proceedings of the 2015 ACM Conference on Special Interest Group on
  Data Communication}, ser. SIGCOMM '15, 2015, pp. 465--478.

\bibitem{CONGA}
M.~Alizadeh, T.~Edsall, S.~Dharmapurikar, R.~Vaidyanathan, K.~Chu,
  A.~Fingerhut, V.~T. Lam, F.~Matus, R.~Pan, N.~Yadav, and G.~Varghese,
  ``Conga: Distributed congestion-aware load balancing for datacenters,'' in
  \emph{Proceedings of the 2014 ACM Conference on SIGCOMM}, ser. SIGCOMM '14,
  2014, pp. 503--514.

\bibitem{dibs}
\BIBentryALTinterwordspacing
K.~Zarifis, R.~Miao, M.~Calder, E.~Katz-Bassett, M.~Yu, and J.~Padhye, ``Dibs:
  Just-in-time congestion mitigation for data centers,'' in \emph{Proceedings
  of the Ninth European Conference on Computer Systems}, ser. EuroSys
  '14.\hskip 1em plus 0.5em minus 0.4em\relax ACM, 2014, pp. 6:1--6:14.
  [Online]. Available: \url{http://doi.acm.org/10.1145/2592798.2592806}
\BIBentrySTDinterwordspacing

\bibitem{rcp}
N.~Dukkipati, M.~Kobayashi, R.~Zhang-Shen, and N.~McKeown, ``Processor sharing
  flows in the internet,'' in \emph{Proceedings of the 13th International
  Conference on Quality of Service}, ser. IWQoS'05.\hskip 1em plus 0.5em minus
  0.4em\relax Springer-Verlag, 2005, pp. 271--285.

\bibitem{numfabric}
K.~Nagaraj, D.~Bharadia, H.~Mao, S.~Chinchali, M.~Alizadeh, and S.~Katti,
  ``Numfabric: Fast and flexible bandwidth allocation in datacenters,'' in
  \emph{Proceedings of the 2016 Conference on ACM SIGCOMM 2016 Conference},
  ser. SIGCOMM '16, 2016, pp. 188--201.

\bibitem{expresspass}
I.~Cho, K.~Jang, and D.~Han, ``Credit-scheduled delay-bounded congestion
  control for datacenters,'' in \emph{Proceedings of the Conference of the ACM
  Special Interest Group on Data Communication}, ser. SIGCOMM '17.\hskip 1em
  plus 0.5em minus 0.4em\relax New York, NY, USA: ACM, 2017, pp. 239--252.

\bibitem{Duato:original}
\BIBentryALTinterwordspacing
J.~Duato, ``A new theory of deadlock-free adaptive routing in wormhole
  networks,'' \emph{IEEE Trans. Parallel Distrib. Syst.}, vol.~4, no.~12, pp.
  1320--1331, Dec. 1993. [Online]. Available:
  \url{http://dx.doi.org/10.1109/71.250114}
\BIBentrySTDinterwordspacing

\bibitem{hot-potato}
P.~Baran, ``On distributed communications networks,'' \emph{Communications
  Systems, IEEE Transactions on}, vol.~12, no.~1, pp. 1--9, March 1964.

\bibitem{tail-cacm}
\BIBentryALTinterwordspacing
J.~Dean and L.~A. Barroso, ``The tail at scale,'' \emph{Commun. ACM}, vol.~56,
  no.~2, pp. 74--80, Feb. 2013. [Online]. Available:
  \url{http://doi.acm.org/10.1145/2408776.2408794}
\BIBentrySTDinterwordspacing

\bibitem{VIA}
\BIBentryALTinterwordspacing
D.~Dunning, G.~Regnier, G.~McAlpine, D.~Cameron, B.~Shubert, F.~Berry, A.~M.
  Merritt, E.~Gronke, and C.~Dodd, ``The virtual interface architecture,''
  \emph{IEEE Micro}, vol.~18, no.~2, pp. 66--76, Mar. 1998. [Online].
  Available: \url{http://dx.doi.org/10.1109/40.671404}
\BIBentrySTDinterwordspacing

\bibitem{ib}
\BIBentryALTinterwordspacing
``Infiniband trade association,'' 2017. [Online]. Available:
  \url{http://www.infinibandta.org}
\BIBentrySTDinterwordspacing

\bibitem{roce}
\BIBentryALTinterwordspacing
``Rdma over converged ethernet,'' 2017, accessed: 2019-12-28. [Online].
  Available: \url{http://www.mellanox.com/page/products_dyn?product_family=79.}
\BIBentrySTDinterwordspacing

\bibitem{agilio}
``Agilio cx smartnics,'' \url{https://www.netronome.com/products/agilio-cx},
  accessed: 2019-12-28.

\bibitem{connectx5}
``Mellanox connectx-5 product brief,''
  \url{http://www.mellanox.com/related-docs/prod_adapter_cards/PB_ConnectX-5_VPI_Card.pdf},
  accessed: 2019-12-28.

\bibitem{mittal-rdma}
R.~Mittal, A.~Shpiner, A.~Panda, E.~Zahavi, A.~Krishnamurthy, S.~Ratnasamy, and
  S.~Shenker, ``Revisiting network support for rdma,'' in \emph{Proceedings of
  the 2018 Conference of the ACM Special Interest Group on Data Communication},
  ser. SIGCOMM '18, 2018, pp. 313--326.

\bibitem{clos}
C.~Clos, ``A study of non-blocking switching networks,'' \emph{Bell Labs
  Technical Journal}, vol.~32, no.~2, pp. 406--424, 1953.

\bibitem{valleyfree}
L.~Gao, ``On inferring autonomous system relationships in the internet,''
  \emph{IEEE/ACM Transactions on Networking}, vol.~9, no.~6, pp. 733--745, Dec
  2001.

\bibitem{duato97}
F.~Silla and J.~Duato, ``Improving the efficiency of adaptive routing in
  networks with irregular topology,'' in \emph{Proceedings Fourth International
  Conference on High-Performance Computing}, Dec 1997, pp. 330--335.

\bibitem{tagger}
S.~{Hu}, Y.~{Zhu}, P.~{Cheng}, C.~{Guo}, K.~{Tan}, J.~{Padhye}, and K.~{Chen},
  ``Tagger: Practical pfc deadlock prevention in data center networks,''
  \emph{IEEE/ACM Transactions on Networking}, vol.~27, no.~2, pp. 889--902,
  April 2019.

\bibitem{Dally-deadlocks}
W.~J. Dally and C.~L. Seitz, ``Deadlock-free message routing in multiprocessor
  interconnection networks,'' \emph{IEEE Transactions on Computers}, vol. C-36,
  no.~5, pp. 547--553, May 1987.

\bibitem{DCdeadlocks}
\BIBentryALTinterwordspacing
S.~Hu, Y.~Zhu, P.~Cheng, C.~Guo, K.~Tan, J.~Padhye, and K.~Chen, ``Deadlocks in
  datacenter networks: Why do they form, and how to avoid them,'' in
  \emph{Proceedings of the 15th ACM Workshop on Hot Topics in Networks}, ser.
  HotNets '16.\hskip 1em plus 0.5em minus 0.4em\relax New York, NY, USA: ACM,
  2016, pp. 92--98. [Online]. Available:
  \url{http://doi.acm.org/10.1145/3005745.3005760}
\BIBentrySTDinterwordspacing

\bibitem{pfabric}
\BIBentryALTinterwordspacing
M.~Alizadeh, S.~Yang, M.~Sharif, S.~Katti, N.~McKeown, B.~Prabhakar, and
  S.~Shenker, ``pfabric: Minimal near-optimal datacenter transport,'' in
  \emph{Proceedings of the ACM SIGCOMM 2013 Conference on SIGCOMM}, ser.
  SIGCOMM '13.\hskip 1em plus 0.5em minus 0.4em\relax New York, NY, USA: ACM,
  2013, pp. 435--446. [Online]. Available:
  \url{http://doi.acm.org/10.1145/2486001.2486031}
\BIBentrySTDinterwordspacing

\bibitem{tcpbolt}
B.~Stephens, A.~Cox, A.~Singla, J.~Carter, C.~Dixon, and W.~Felter, ``Practical
  dcb for improved data center networks,'' in \emph{INFOCOM, 2014 Proceedings
  IEEE}, April 2014, pp. 1824--1832.

\bibitem{rcp-router}
M.~Flores, A.~Wenzel, and A.~Kuzmanovic, ``Enabling router-assisted congestion
  control on the internet,'' in \emph{2016 IEEE 24th International Conference
  on Network Protocols (ICNP)}, Nov 2016, pp. 1--10.

\bibitem{nandita-thesis}
N.~Dukkipati, ``Rate control protocol (rcp): Congestion control to make flows
  complete quickly,'' Ph.D. dissertation, Department of Electrical Engineering,
  Stanford University, Stanford, CA, USA, 2008, aAI3292347.

\bibitem{d2tcp}
B.~Vamanan, J.~Hasan, and T.~Vijaykumar, ``Deadline-aware datacenter tcp
  (d2tcp),'' in \emph{Proceedings of the ACM SIGCOMM 2012 Conference on
  Applications, Technologies, Architectures, and Protocols for Computer
  Communication}, ser. SIGCOMM '12, 2012.

\bibitem{ICTCP}
\BIBentryALTinterwordspacing
H.~Wu, Z.~Feng, C.~Guo, and Y.~Zhang, ``Ictcp: Incast congestion control for
  tcp in data center networks,'' in \emph{Proceedings of the 6th International
  COnference}, ser. Co-NEXT '10.\hskip 1em plus 0.5em minus 0.4em\relax New
  York, NY, USA: ACM, 2010, pp. 13:1--13:12. [Online]. Available:
  \url{http://doi.acm.org/10.1145/1921168.1921186}
\BIBentrySTDinterwordspacing

\bibitem{d3}
\BIBentryALTinterwordspacing
C.~Wilson, H.~Ballani, T.~Karagiannis, and A.~Rowtron, ``Better never than
  late: meeting deadlines in datacenter networks,'' in \emph{Proceedings of the
  ACM SIGCOMM 2011 conference}, ser. SIGCOMM '11.\hskip 1em plus 0.5em minus
  0.4em\relax New York, NY, USA: ACM, 2011, pp. 50--61. [Online]. Available:
  \url{http://doi.acm.org/10.1145/2018436.2018443}
\BIBentrySTDinterwordspacing

\bibitem{pdq}
\BIBentryALTinterwordspacing
C.-Y. Hong, M.~Caesar, and P.~B. Godfrey, ``Finishing flows quickly with
  preemptive scheduling,'' in \emph{Proceedings of the ACM SIGCOMM 2012
  conference on Applications, technologies, architectures, and protocols for
  computer communication}, ser. SIGCOMM '12.\hskip 1em plus 0.5em minus
  0.4em\relax New York, NY, USA: ACM, 2012, pp. 127--138. [Online]. Available:
  \url{http://doi.acm.org/10.1145/2342356.2342389}
\BIBentrySTDinterwordspacing

\bibitem{karuna}
L.~Chen, K.~Chen, W.~Bai, and M.~Alizadeh, ``Scheduling mix-flows in commodity
  datacenters with karuna,'' in \emph{Proceedings of the 2016 Conference on ACM
  SIGCOMM 2016 Conference}, ser. SIGCOMM '16, 2016, pp. 174--187.

\bibitem{ups}
R.~Mittal, R.~Agarwal, S.~Ratnasamy, and S.~Shenker, ``Universal packet
  scheduling,'' in \emph{Proceedings of the 13th Usenix Conference on Networked
  Systems Design and Implementation}, ser. NSDI'16, 2016, pp. 501--521.

\bibitem{phost}
P.~X. Gao, A.~Narayan, G.~Kumar, R.~Agarwal, S.~Ratnasamy, and S.~Shenker,
  ``phost: Distributed near-optimal datacenter transport over commodity network
  fabric,'' in \emph{Proceedings of the 11th ACM Conference on Emerging
  Networking Experiments and Technologies}, ser. CoNEXT '15, 2015, pp.
  1:1--1:12.

\bibitem{qcn}
\BIBentryALTinterwordspacing
``Qcn: Quantized congestion notification an overview,'' 2007. [Online].
  Available: \url{http://www.ieee802.org/1/files/public/docs2007/au_
  prabhakar_qcn_overview_geneva.pdf}
\BIBentrySTDinterwordspacing

\bibitem{diatcp}
J.-H. Hwang, J.~S. Yoo, and N.~Choi, ``Deadline and incast aware tcp for cloud
  data center networks,'' \emph{Computer Networks}, vol.~68, pp. 20--34, 08
  2014.

\bibitem{MPTCPNSDI12}
C.~Raiciu, C.~Paasch, S.~Barr{\`E}, A.~Ford, M.~Honda, F.~Duchene,
  O.~Bonaventure, and M.~Handley, ``How hard can it be? designing and
  implementing a deployable multipath tcp,'' in \emph{USENIX Symposium of
  Networked Systems Design and Implementation (NSDI'12), San Jose (CA)}, 2012.

\bibitem{hermes-sigcomm17}
H.~Zhang, J.~Zhang, W.~Bai, K.~Chen, and M.~Chowdhury, ``Resilient datacenter
  load balancing in the wild,'' in \emph{Proceedings of the Conference of the
  ACM Special Interest Group on Data Communication}, ser. SIGCOMM '17, 2017,
  pp. 253--266.

\bibitem{drill-sigcomm17}
S.~Ghorbani, Z.~Yang, P.~B. Godfrey, Y.~Ganjali, and A.~Firoozshahian, ``Drill:
  Micro load balancing for low-latency data center networks,'' in
  \emph{Proceedings of the Conference of the ACM Special Interest Group on Data
  Communication}, ser. SIGCOMM '17, 2017, pp. 225--238.

\end{thebibliography}

\begin{IEEEbiography}[{\includegraphics[width=1in,height=1.25in,clip,keepaspectratio]
{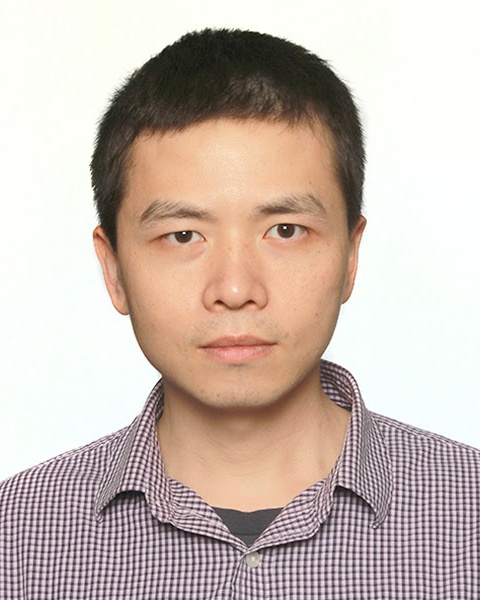}}]{Jiachen Xue} received a Ph.D. in Computer Engineering from Purdue University (2017), an M.S in Electrical Engineering (2008) from Arizona State University, and a bachelor in Computer Engineering (2006) from Beihang University. He currently works as a senior distributed system engineer at NVIDIA.
\end{IEEEbiography}

% if you will not have a photo at all:
\begin{IEEEbiography}[{\includegraphics[width=1in,height=1.25in,clip,keepaspectratio]
{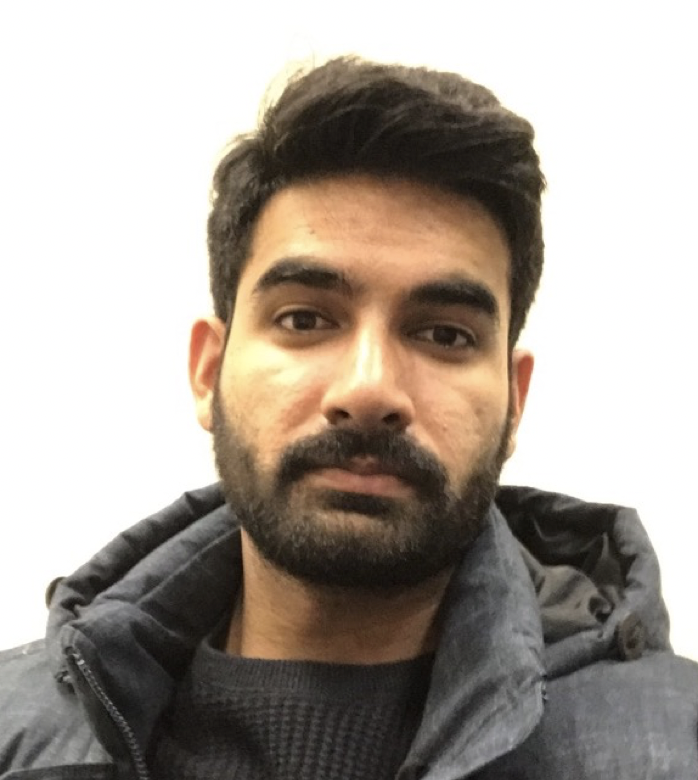}}]{Muhammad Usama Chaudhry} received a M.S. in Computer Science Department at the University of Illinois at Chicago (2019), 
and a bachelors degrees from the National University of Computer Science, Islamabad (2014). He currently works as a softwate engineer at VMWare 
Inc. 
\end{IEEEbiography}

%\newpage

\begin{IEEEbiography}[{\includegraphics[width=1in,height=1.25in,clip,keepaspectratio]
{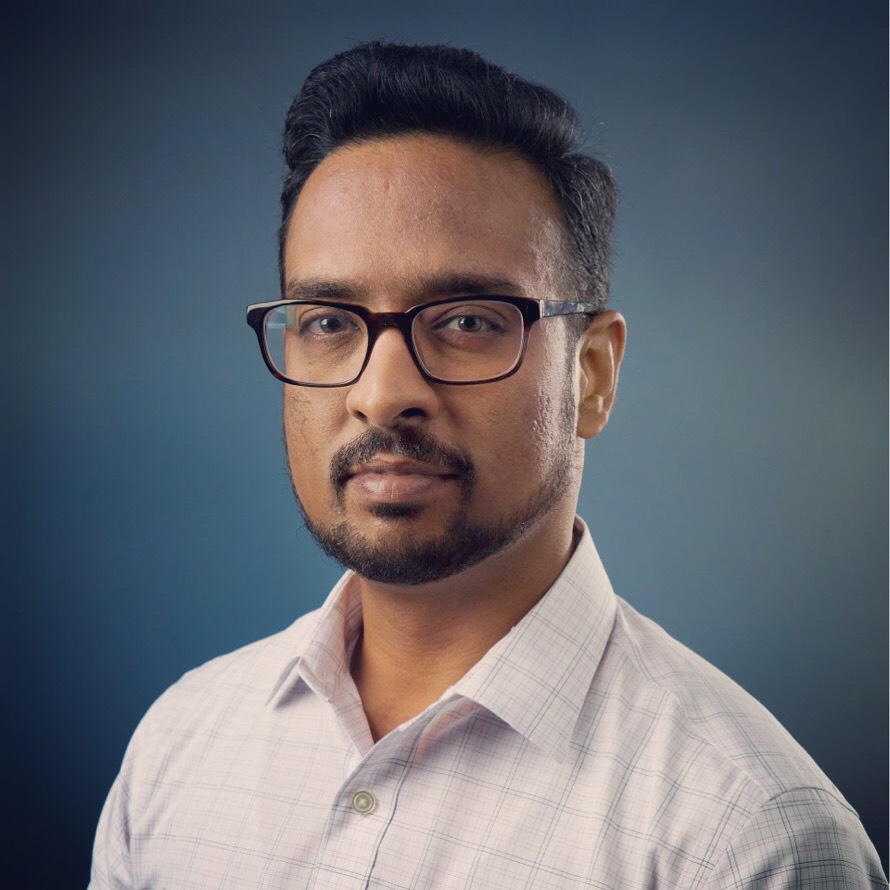}}]{Balajee Vamanan}
is an Assistant Professor in the Department of Computer Science at University of Illinois at Chicago (UIC). His research interests span various aspects of computer networks and computer systems. He received his Ph.D. from Purdue University in 2015. Prior to his graduate study, he worked in NVIDIA as a design engineer. 
\end{IEEEbiography}

\begin{IEEEbiography}[{\includegraphics[width=1in,height=1.25in,clip,keepaspectratio]
{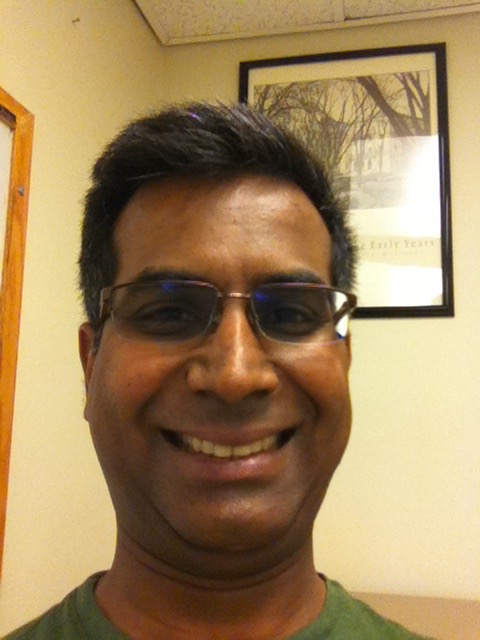}}]{T. N. Vijaykumar}
is a Professor in the School of Electrical and Computer Engineering at Purdue University.
His research interests are in computer architecture targeting various aspects of performance, power, programmability,  and reliability of computer hardware and systems.
Recognition of his work include a 1999 NSF CAREER Award, IEEE Micro's Top Picks from 2003 and 2005 computer architecture papers, listing in the International Symposium on Computer Architecture (ISCA) Hall of Fame, and the  first prize in the 2009 Burton D. Morgan Business Plan Competition.
\end{IEEEbiography}

\begin{IEEEbiography} [{\includegraphics[width=1in,height=1.25in,clip,keepaspectratio]
{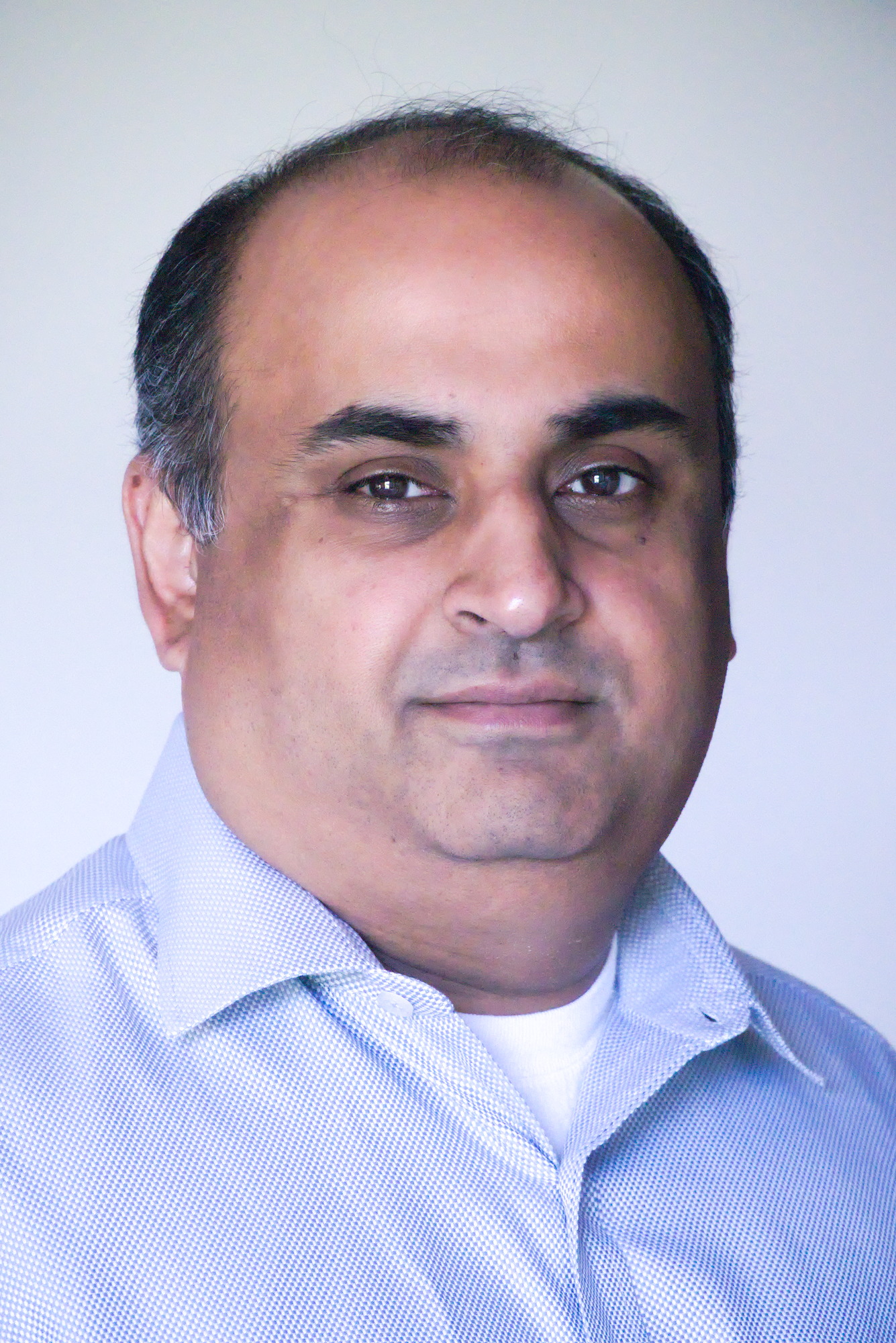}}]{Mithuna Thottethodi}
is an Associate Professor of Electrical and Computer Engineering at Purdue University. His research interests include computer architecture, distributed systems, and networks. He received a B.Tech. (Hons) degree in Computer Science and Engineering from the Indian Institute of Technology, Kharagpur and his Ph.D. in Computer Science from Duke University. He received the NSF CAREER award in 2007.
\end{IEEEbiography}

\end{document}